\documentclass[onecolumn,12pt]{article}
\usepackage{times,comment,overpic,graphicx,amssymb,amsmath,xcolor,multicol,overpic,colortbl,tabularx,tabu,multicol,xr,authblk,abstract,float,multirow,array,booktabs,forloop}
\usepackage[disable]{todonotes}
\usepackage{url}

\usepackage{scicite}
\newenvironment{sciabstract}{%
\begin{quote} \bf}
{\end{quote}}

\makeatletter
\newcommand*{\addFileDependency}[1]{
  \typeout{(#1)}
  \@addtofilelist{#1}
  \IfFileExists{#1}{}{\typeout{No file #1.}}
}
\makeatother
 
\newcommand*{\myexternaldocument}[1]{%
    \externaldocument{#1}%
    \addFileDependency{#1.tex}%
    \addFileDependency{#1.aux}%
}
\myexternaldocument{SI}

\definecolor{myRed}{rgb}{1,0.8,0.8}
\definecolor{rev}{rgb}{0.9019, 0.0274,  0.1647058}
\definecolor{myGreen}{rgb}{0.4274,   0.7529,   0.28235}
\definecolor{myBlue}{rgb}{0.2588,   0.3098,   0.643137}
\definecolor{myCyan}{rgb}{0.46568627,  0.76372549,  0.81960784}
\definecolor{myMagenta}{rgb}{0.70588,   0.29019,   0.61960}
\definecolor{myYellow}{rgb}{1,1,.1}

\title{AI exposure predicts unemployment risk}

\author[1,2,3,4,*]{Morgan R. Frank}
\author[4,5,6]{Yong-Yeol Ahn}
\author[3,7,8]{Esteban Moro}

\affil[1]{Department of Informatics and Networked Systems, University of Pittsburgh, Pittsburgh, PA 15216 USA}
\affil[2]{Digital Economy Lab, Institute for Human-Centered Artificial Intelligence, Stanford University, Stanford, CA 94305 USA}
\affil[3]{Media Laboratory, Massachusetts Institute of Technology, Cambridge, MA, 02139 USA}
\affil[4]{Connection Science, Massachusetts Institute of Technology, Cambridge, MA, USA}
\affil[5]{Center for Complex Networks and Systems Research, Luddy School of Informatics,
Computing, and Engineering, Indiana University, Bloomington, IN, USA}
\affil[6]{Indiana University Network Science Institute, Indiana University, Bloomington, IN, USA}
\affil[7]{Department of Mathematics \& GISC, Universidad Carlos III de Madrid, 28911 Leganes, Spain}
\affil[8]{Institute for Data, Systems, and Society, Massachusetts Institute of Technology, Cambridge, MA, USA}
\affil[*]{To whom correspondence should be addressed. E-mail: mrfrank@pitt.edu}

\date{}
\begin{document}

\maketitle

\begin{sciabstract}
    Is artificial intelligence (AI) disrupting jobs and creating unemployment?
    Despite many attempts to quantify occupations' exposure to AI, inconsistent validation obfuscates the relative benefits of each approach.
    A lack of disaggregated labor outcome data, including unemployment data, further exacerbates the issue.
    Here, we assess which models of AI exposure predict job separations and unemployment risk using new occupation-level unemployment data by occupation from each US state's unemployment insurance office spanning 2010 through 2020.
    Although these AI exposure scores have been used by governments and industry, we find that individual AI exposure models are not predictive of unemployment rates, unemployment risk, or job separation rates.
    However, an ensemble of those models exhibits substantial predictive power suggesting that competing models may capture different aspects of AI exposure that collectively account for AI's variable impact across occupations, regions, and time.
    Our results also call for dynamic, context-aware, and validated methods for assessing AI exposure.
    Interactive visualizations for this study are available at \url{https://sites.pitt.edu/~mrfrank/uiRiskDemo/}. 
\end{sciabstract}

\section*{Introduction}
Artificial intelligence (AI)---broadly defined as technology that learns patterns from data---has revived fears of technological unemployment~\cite{toole2020inventing,mcclure2018you,pew2017automation,keynes2010economic,leontief1952machines}.
Some argue that AI is similar to previous technologies that did not result in massive unemployment~\cite{aghion1990model,Aghion,bartelsman2004microeconomic}.
But others argue that AI is a different class of technology because it can perform cognitive work and thus warrants new consideration.
This has prompted efforts to estimate occupation-level AI exposure~\cite{frey2017the,arntz2016risk,brynjolfsson2018what,felten2018method,webb2019impact,tolan2021measuring} that can inform both economic research and policy-making to support workers' adaption to the future of work.
For example, the recent Biden Administration Blueprint for an AI Bill of Rights~\cite{aibill} calls for ``training, assessment, and oversight to combat automation bias.''  
To achieve these goals, policymakers and workers need validated measures for AI exposure~\cite{frank2019toward,blien2022contradictory,brynjolfsson2018artificial,brynjolfsson2014second,mcclure2018you}.

However, validation is challenging because of the diverse outcomes that AI exposure may produce and interactions between various occupations, locations, and time.
Existing studies have compared AI exposure scores to employment or wage losses~\cite{acemoglu2020ai,fossen2022new} with mixed results.
Government and industry analyses, including from the Brookings Institution~\cite{muro2017digitalization}, the US Bureau of Labor Statistics (BLS)~\cite{mean2023job} and the Organisation for Economic Co-operation and Development (OECD)~\cite{nedelkoska2018automation,georgieff2021happened,lane2021impact,georgieff2021artificial}, use AI exposure scores to analyze workers' exposure to technology in order to predict potential \emph{job losses}. 
However, recent reports~\cite{handel2022growth,dell2020assessing,mean2023job} from the US BLS find no noticeable decrease in \emph{employment} for ``high-exposure'' occupations and suggest that new data is needed to understand the labor effects of technology.

These mixed results may arise from focusing on the wrong symptoms of AI exposure.
For example, \emph{employment} data may be uncorrelated with \emph{job losses}.
Unemployment or job losses for an occupation can increase or decrease even as employment for that occupation increases~\cite{mean2023job}. 
Technology alters an occupation's required workplace activities (i.e., within-occupation skill change), which can displace workers who cannot adapt (i.e., job separations from employer firing or worker quitting)~\cite{mortensen1998technological,jovanovic1979job,fossen2022new}.
Finally, if displaced workers cannot find new employment quickly, they may rely on unemployment benefits while they continue job seeking.
Thus, only examining detailed employment data would overlook other major outcomes of AI exposure~\cite{georgieff2021happened}, including skill change~\cite{mean2023job}, job separations, and unemployment.
Even when unemployment is occurring, employment and wages may remain stable potentially leading researchers to erroneously conclude that no detrimental outcomes had occurred
For instance, new workers may replace other workers if they can complement new technology, thus boosting employment and/or wages for the exposed occupation even though a disruption occurred (e.g., bank tellers and ATMS~\cite{bessen2015computer,bessen2018ai}).

Quantifying all of the consequences from AI exposure requires more than employment and/or wage data.
However, such data is typically available only on aggregate (e.g., total unemployment or total job separations using LAUS from the US Bureau of Labor Statistics (BLS)).
For example, both total employment and total unemployment can not elucidate all labor dynamics of specific occupations by themselves.
Breaking this barrier requires \emph{high-resolution} unemployment data that can be broken down by time, place, and occupation.

In this article, we break this barrier to detailed unemployment data by building a high-resolution dataset from monthly unemployment counts by most-recent occupation from each state's unemployment insurance office.
After describing occupation AI exposure scores from existing studies, we see which scores, if any, predict an increased probability of receiving unemployment benefits (which we call \emph{unemployment risk}), job separations (i.e., job quitting or firing), or within-occupation skill change using multiple regression analysis.
Although individual scores are not predictive of unemployment risk, job separations, or skill change, an ensemble model is predictive even after controlling for regional fixed effects, temporal effects, and occupations' skill requirements.
We conclude by analyzing which AI exposure scores are most applicable in different parts of the US economy (e.g., by state or occupation) and observe strong geographical heterogeneity in the applicability of each AI exposure score suggesting that efforts to predict AI job losses cannot rely on any one score.
Combined, these results demonstrate that employment and wage data may miss other detrimental labor dynamics from AI technology and that efforts using only one AI exposure score will misrepresent AI's impact on the future of work.

\begin{table}[t]
    \centering\scriptsize
    \def\arraystretch{1.3}
    \begin{tabularx}{\textwidth}{
    >{\raggedright\hsize=.04\hsize}X|
    >{\hsize=.12\hsize}X
    >{\hsize=.09\hsize}X
    >{\hsize=.3\hsize}X
    >{\hsize=.44\hsize}X
       }
        Wave & Study & Year First Available & Scores & Description \\ \hline
        1 & O*NET Bachelors & 2003 & (denoted {\bf \%college})
        &The fraction of workers in an occupation with a bachelor's degree. \\ 
        & Acemoglu \& Autor~\cite{acemoglu2011skills} & 2011 & Computer Usage ({\bf Comp.Use}), Routine Cognitive ({\bf R.Cog.}), Routine Manual ({\bf R.Man.}) 
        & Assess occupations on computer usage, routineness, and cognitive or manual requirements. 
         \\\hline 
        2 & Frey \& Osborne~\cite{frey2017the} & preprint 2013 & Probability of Computerization ({\bf auto})
        &Combined a subset of occupation skills with subjective assessments of fully automatable or non-automatable occupations. \\
        & Arntz et al~\cite{arntz2016risk} & 2016 & Probability of Computerization ({\bf auto2})
        & Considered a complete set of occupations' skills to assess automation risk in OECD countires. \\
        & O*NET Degree of Automation & 2016 & ({\bf Deg.Auto.})
        & The relative amounts of routine versus challenging work the worker will perform as part of a job. \\\hline
        3 & Brynjolfsson et al~\cite{brynjolfsson2018what} & 2018 & Suitability for Machine Learning ({\bf SML})
        & Surveyed ML experts in order to assess occupations' task suitability for ML. \\ 
        & Felten et al~\cite{felten2018method} & 2018 & ({\bf AI2})
        & Crowdsource gig workers to establish connections between AI application capabilities and occupation abilities. \\ 
        & Webb~\cite{webb2019impact} & 2019 & \% AI Exposure ({\bf AI}), \% Software Exposure ({\bf Software}), \% Robot Exposure ({\bf Robot})
        &Uses NLP to compare technology patents to occupation tasks. \\ \hline
    \end{tabularx}
    \caption{
        Waves of studies estimating AI exposure by occupation.
        Methodologies have evolved from solely theoretical motivations (Wave 1) to greater specificity into occupations' skills (Wave 2) to connecting skills to the capabilities of specific technologies (Wave 3).
        Scores are taken from each study; short-hands for each score are provided in parentheses. 
    }
    \label{tbl:studies}
\end{table}

\section*{Quantifying Unemployment Risk by Occupation}

We collect data from each US state's unemployment benefits office detailing the US unemployment benefit recipients by state, month, and their most recent occupation.
Summing unemployment counts across occupations in a given state and month will correspond to the total unemployment reported by BLS LAUS.
However, unlike LAUS total unemployment, this data is stratified by occupation, location, and time and thus offers better resolution with which to study how labor disruptions differentially impact workers by occupation and labor market.
Combined with employment statistics from the US BLS, this data enables a high-resolution measure of unemployment pressure $p(u|soc,s,t)$---namely the unemployment risk of an occupation $soc$ given location (i.e., state $s$) and time (i.e., year and month $t$).
We explain this methodology with more detail in SI Section 3.
Unemployment risk varies by occupation (e.g., construction workers versus transportation workers) while controlling for states' total labor market size and occupations' local employment share (e.g., more retail workers may receive unemployment benefits in a state where retail workers are a larger share of employment).
Data from the BLS's Job Openings and Labor Turnover Survey (JOLTS) and O*NET database provide data on job separations and within-occupation changes to skill demands.
Taken together, our dataset offers a new opportunity to study the comprehensive set of possible labor outcomes resulting from exposure to technology across skill change, job separations, and unemployment.

We use unemployment risk to test the predictive power of existing models of AI exposure, some of which have been used in policymaking (e.g., the OECD~\cite{georgieff2021happened,georgieff2021artificial}).
Models for occupations' exposure to AI and/or robotics have emerged over the last decade while emphasizing different technologies and industries.
But it remains unclear which labor markets (e.g., states) or occupations are best predicted by each approach, and how to combine their relative strengths.
Here, we will focus on AI exposure while leaving exposure to robotics mostly as future work.
Motivated by the skill-biased technological change theory~\cite{hicks1963theory,violante2008skill,acemoglu2011skills}, a first wave of studies argued that college-educated, high-skill workers, who perform cognitive tasks, are complemented by technology while low-skill workers, who perform manual tasks, are more frequently substituted by technologies like robotics~\cite{autor2008trends,acemoglu2002technical}.
A second wave of studies used BLS O*NET data to model occupations based on their skill requirements~\cite{frey2017the,arntz2016risk}.
Most recently, a third wave of studies compares technological capabilities to job descriptions (i.e., a task-based approach~\cite{acemoglu2021tasks}) by surveying machine learning (ML) experts~\cite{brynjolfsson2018what}, surveying gig workers~\cite{felten2018method}, or applying natural language processing to technology patents~\cite{webb2019impact}.
Table~\ref{tbl:studies} summarizes these studies across all three waves (see SI Section 1 for more information).
These studies span the previous decade and each raise the issue of technology reshaping US labor which suggests that enough time has passed for some impact to have occurred.
While theory and methodology vary across these studies, none of them provide empirical causal analysis to validate the role of technology in labor outcomes.

\section*{Automation Estimates and Unemployment Risk}

\begin{figure}[p]
    \centering
    \includegraphics[width=.99\textwidth]{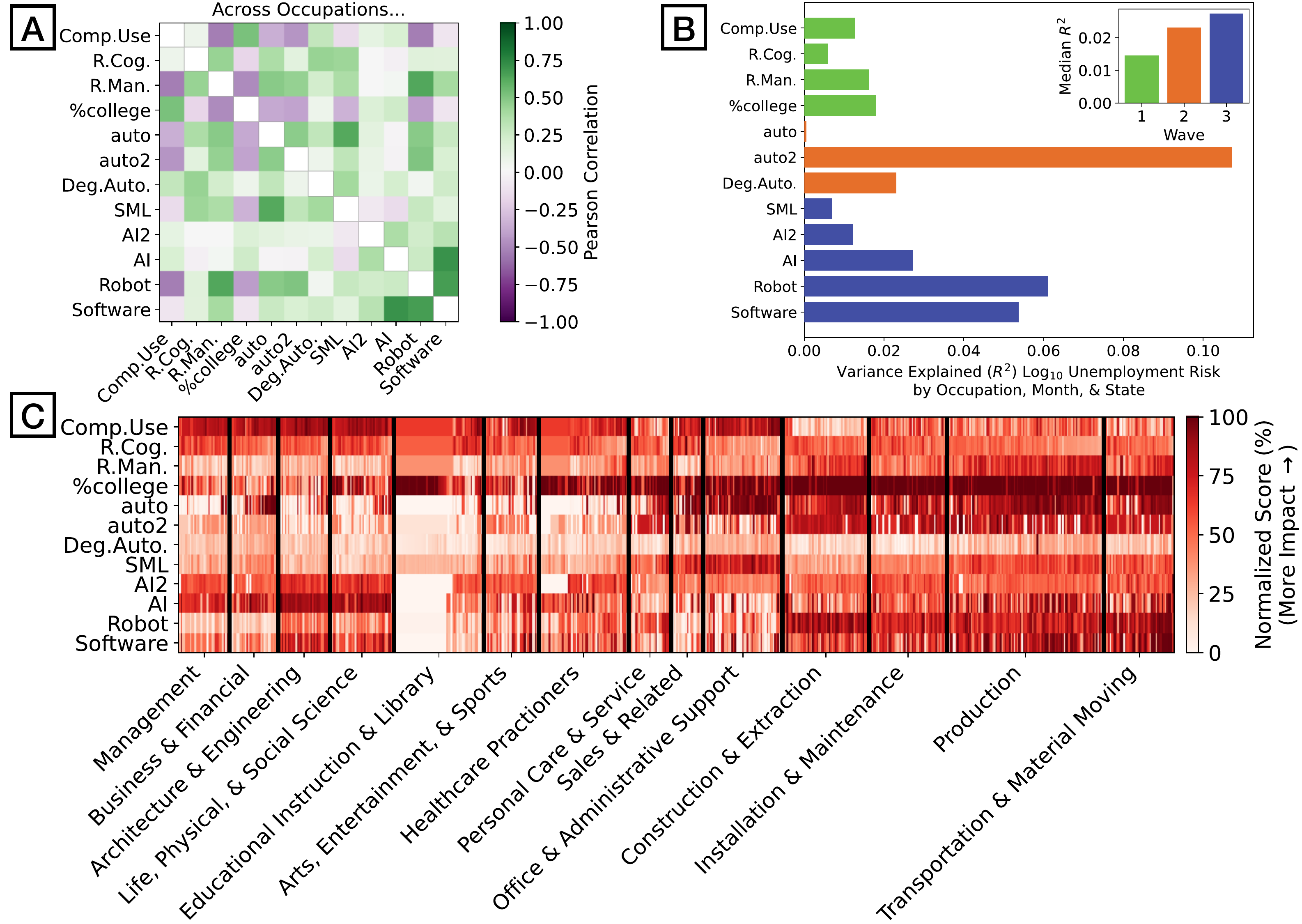}
    \caption{
        Technological exposure scores are not consistent with each other and cannot individually predict unemployment risk well.
        {\bf (A)} The Pearson correlation of pairs of AI exposure scores across occupations.
        Scores are ordered according to the study's Wave (see Table~\ref{tbl:studies}).
        {\bf (B)} 
        The variation in unemployment risk explained by each AI exposure score.
        Colors indicate the score's Wave.
        The inset reports the median variance explained ($R^2$) for scores by Wave.
        {\bf (C)} A heat map detailing occupations' AI exposure scores (color).
        Occupations are grouped according to their major occupation (x-axis).
    }
    \label{fig:scores}
\end{figure}

Since each study shares the common goal of estimating AI exposure, we expect that their estimates will mostly agree with each other.
However, we find that individual exposure scores across studies are not strongly correlated (see Fig.~\ref{fig:scores}A) and can even be anti-correlated when compared across occupations.
For example, Webb's measure for Software usage is negatively correlated with Acemoglu \& Autor's measure of Computer Usage and negatively correlated with the national fraction of workers with a bachelor's degree.
The strongest correlation among individual scores is between AI exposure and Software exposure scores from the same study~\cite{webb2019impact} ($R^2=0.494$), but the variance explained from one score to the next is typically small (average $R^2=0.111$ and median $R^2=0.068$).
This disagreement among exposure scores persists across individual occupations and is present across all occupation categories (see Fig.~\ref{fig:scores}C).

Do any AI exposure scores predict unemployment risk and other labor outcomes?
If so, then technology may be detrimental to some US workers, and comparing different AI exposure scores will reveal a clear winner that policymakers and workers can use.
But, if not, then several possible scenarios arise.
On the one hand, it may be that technology is not a major factor and does not shape labor outcomes.
In this case, no reasonable effort to measure AI exposure would predict unemployment risk or job separations.
On the other hand, if technology indeed impacts workers, then the disagreement among exposure scores (see Fig.~\ref{fig:scores}A\&C) may indicate that researchers have yet to converge on good measures for AI exposure.
Another possibility is that each exposure score describes different types of AI exposure or performs well in different regions; in this case, individual scores might predict only small shares of unemployment risk individually but combine to explain a larger share.

Our analysis supports the last scenario.
First, individual exposure scores do not strongly predict unemployment risk (see Fig.~\ref{fig:scores}B).
For example, occupations' Computer Usage~\cite{acemoglu2011skills} explains less than 2\% of the variation in unemployment risk ($R^2=0.013$) across occupations, states, and time.
The national share of workers of an occupation with a bachelor's degree explains yields $R^2=0.018$.
The most predictive model comes from Arntz et al.~\cite{arntz2016risk} ($R^2=0.107$), which was a response to another Wave 2 study from Frey and Osborne~\cite{frey2017the} ($R^2=0.001$).
On aggregate, the median variance explained increased across the three waves of exposure scores (see Fig.~\ref{fig:scores}B inset) from $R^2=0.015$ for Wave 1 to $R^2=0.027$ for Wave 3.
The most predictive score explains only 10.7\% of variation in unemployment risk across occupations, states, and months. 

\begin{figure}[p]
    \centering
    \includegraphics[width=\textwidth]{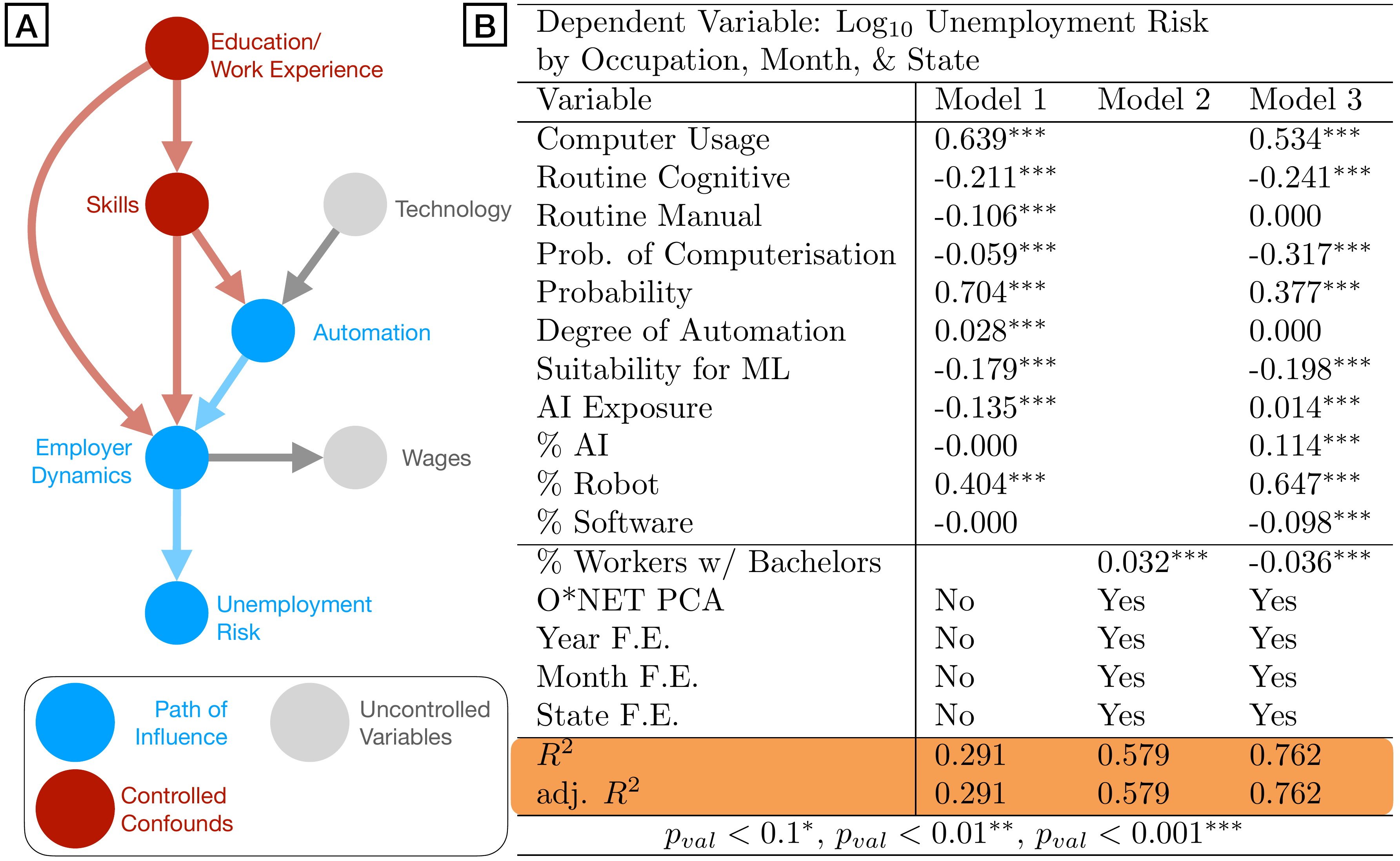}
    \caption{
        Combining AI exposure scores from all studies substantially improves predictions of unemployment risk.
        {\bf (A)} A schematic for the theoretical influence among education, skills, technology, and unemployment risk.
        {\bf (B)} Combined into a single linear model, AI exposure scores capture 29.1\% of the variation in unemployment risk (Model 1).
        Compared to a baseline model with controls for year, seasonality, state, occupations' educational requirements and occupations' O*NET skill requirements (Model 2), the combined model accounts for 76.2\% variation in unemployment risk (Model 3).
        All variables were centered and standardized before LASSO regression.
        Unemployment risk is calculated using monthly data on unemployment recipients from each US state's unemployment insurance office ($N=140,274$).
    }
    \label{fig:combined}
\end{figure}

Yet, taken together as an ensemble, AI exposure scores predict much more variance in unemployment risk across occupations, states, and time.
A single linear model combining all exposure scores accounts for 29.1\% of the variation in unemployment risk (see Fig.~\ref{fig:combined}B Model 1). 
However, this simple approach misses several potential confounds which limits the causality that can be inferred.
For example, a worker's skills or education may determine both their AI exposure and employer's likelihood to hire them.
Or, automation may vary by state and/or time depending on the local economy's industrial composition.
For example, production line robotics would impact blue-collar rural economies more than white-collar urban economies~\cite{frank2018small}.
Figure~\ref{fig:combined}A provides a schematic detailing how additional factors might shape technology's impact on unemployment risk.
Note that wages need not be controlled for in our analysis because wages are determined independently from unemployment risk; however, both factors are shaped by employer dynamics during the hiring process and so we expect them to be correlated (e.g., see SI Section 5).

\begin{figure}[p]
    \centering
    \includegraphics[width=\textwidth]{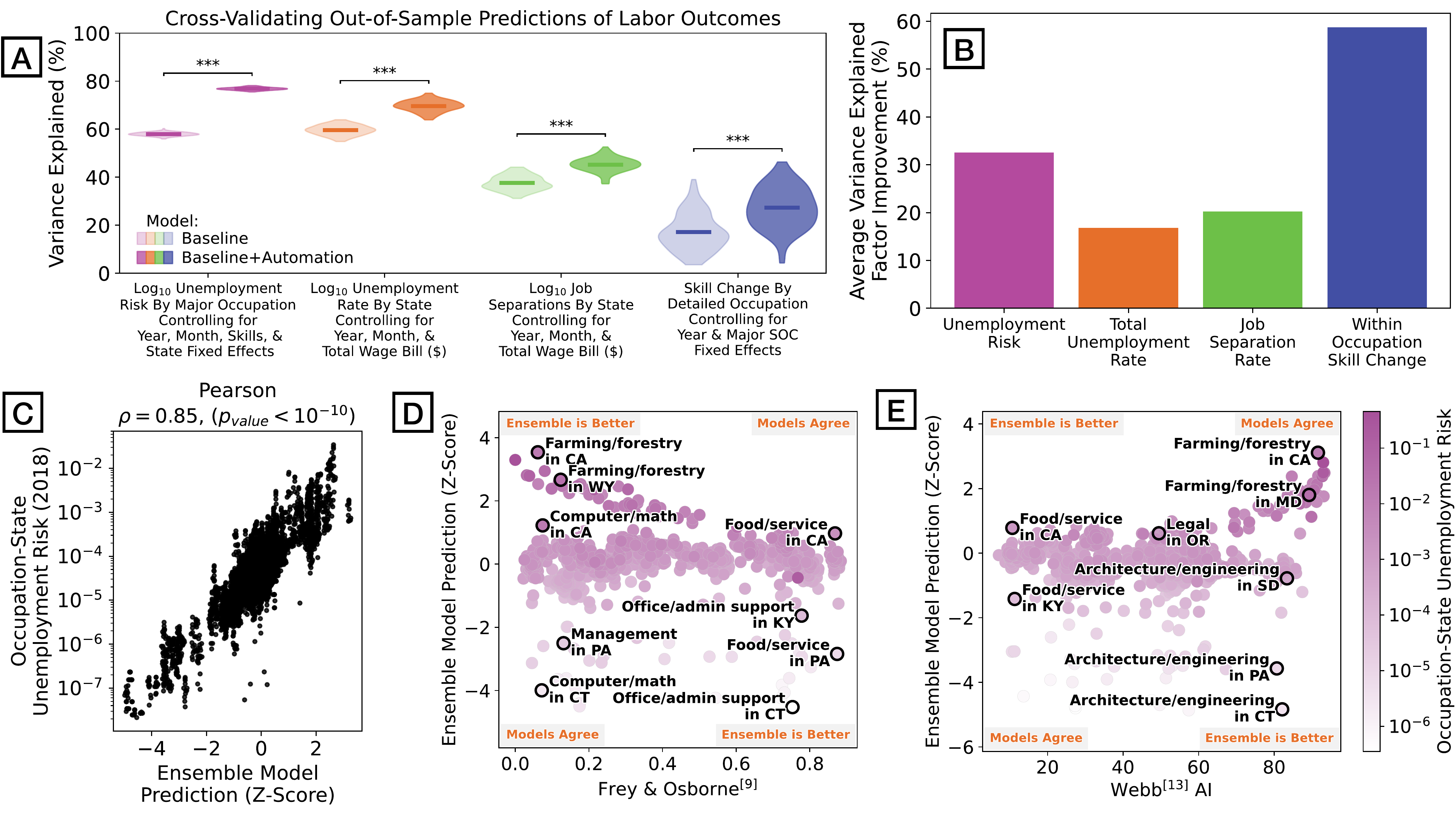}
    \caption{
        Combining AI exposure scores from all studies improves predictions of labor outcomes and reveals shortcomings in individual scores.
        {\bf (A)} Comparing out-of-sample variance explained with and without AI exposure scores.
        Each analysis includes ten independent trials of ten-fold cross-validation training a LASSO regression model; this produces a distribution of 100 observations of each model's predictive performance.
        Horizontal lines represent the average of each distribution.
        For each labor outcome, including the combined AI exposure model significantly improves predictive performance (i.e., two-sample t-test $p$-value $< 10^{-3}$).
        {\bf (B)} The average factor improvement in variance explained for each labor outcome when the combined AI exposure model is added to the baseline model.
        {\bf (C)} Ensemble model's predictions for unemployment risk by occupation and state in 2018 are strongly correlated to actual unemployment risk.
        Monthly unemployment data was averaged to produce annual estimates which were used to calculate unemployment risk.
        {\bf (D)} Compared to the ensemble model (y-axis), predictions from Frey \& Osborne's~\cite{frey2017the} (x-axis) underestimate unemployment risk (color) for Computer/math occupations in California and overestimate unemployment risk for Maintenance occupations in Connecticut using 2013 data.
        {\bf (E)} Compared to the ensemble model, predictions using Webb's AI score~\cite{webb2019impact} underestimate unemployment risk for Food prep/service occupations in California and overestimate unemployment risk for Architecture/engineering occupations in South Dakota using 2019 data.
    }
    \label{fig:crossValidation}
\end{figure}

While other potential confounds may exist, we expect that many of these factors would affect the correlation between technology and unemployment through skills or education.
This observation informs our baseline model, which controls for occupations' O*NET skill requirements (i.e., using principal component analysis. See SI Section4 for methodology), the national fraction of workers in each occupation with a bachelor's degree, and fixed effects for year, month, and state.
This baseline model accounts for 57.9\% of variation in unemployment risk across occupations.
But, adding AI exposure scores to the baseline model accounts for an additional 18.3 percentage points of variation in unemployment risk (i.e., 76.2\% variation explained. See Fig.~\ref{fig:combined}B Models 2 \& 3).
We observe similar performance gains using out-of-sample cross-validation (see Fig.~\ref{fig:crossValidation}A).
We provide full regression tables in SI Section 5.
A large proportion of variation explained indicates that many sources of unemployment risk affect workers through education, workers' skills, seasonality, and regional factors, but the additional 18.3\% of variation explained with the inclusion of the ensemble model (i.e., Model 3) suggests substantial influence from technology.

How do the different AI exposure scores contribute to the ensemble models predictions?
The ensemble model is differentially leveraging the strengths of each model to produce better predictions (e.g., see Fig.~\ref{fig:crossValidation}C).
Comparing individual scores to the ensemble model's predictions highlights the states and occupations where that methodology underperforms.
For example, predictions from Frey \& Osborne’s~\cite{frey2017the} underestimate unemployment risk for Computer/math occupations in California and overestimate unemployment risk for Office/Administrative occupations in Connecticut in 2013 (i.e., the year that scores became available. See Fig.~\ref{fig:crossValidation}D).
As another example, predictions using Webb's AI score~\cite{webb2019impact} underestimate unemployment risk for Food prep/service occupations in California and overestimate unemployment risk for Architecture/engineering occupations in Pennsylvania using 2019 data compared to the predictions of the ensemble model (see Fig.~\ref{fig:crossValidation}E).

\section*{Technology, Job Separations, and Changing Skill Demands}

\begin{figure}[p]
    \centering
    \includegraphics[width=\textwidth]{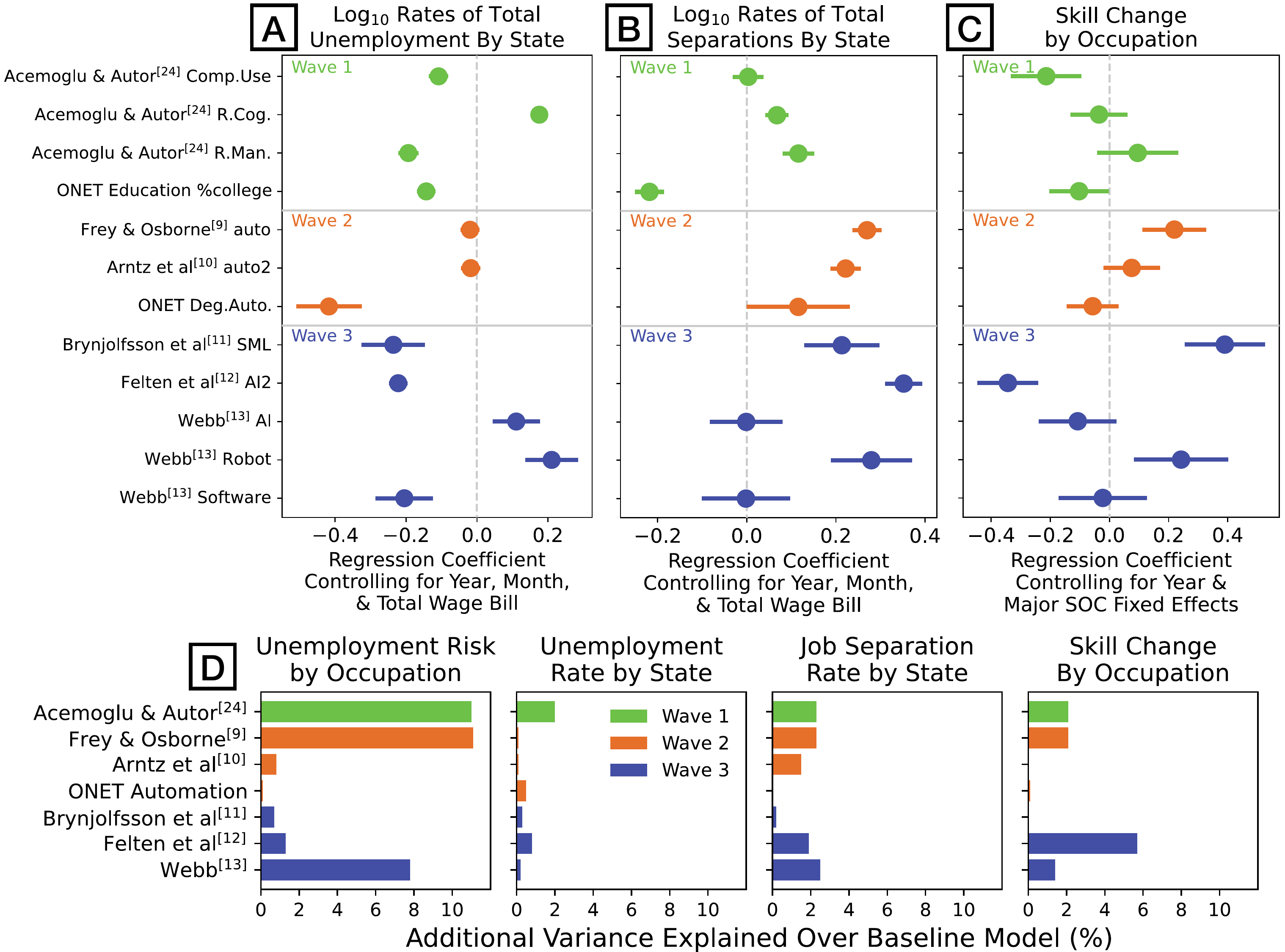}
    \caption{
        AI exposure scores individually predict small amounts of variation states' total unemployment, states' total job separations, and within occupation skill change.
        Linear regression coefficients reporting the relationship between each exposure scores and {\bf (A)} total unemployment rate, {\bf (B)} total job separation rates, and {\bf (C)} within-occupation skill change.
        We provide 95\% confidence intervals.
        Colors represent study waves.
        {\bf (D)} For each labor outcome, the additional predictive performance (i.e., adjusted $R^2$) when exposure scores are added to the baseline model.
        In all plots, all variables were centered and standardized.
        Full regression tables are available in the SI.
    }
    \label{fig:individual}
\end{figure}

What about other labor outcomes?
For example, job separations may occur without contributing to unemployment if displaced workers quickly find new employment~\cite{acemoglu2019automation}.
Similar to unemployment risk, depending on the score used, correlations varied between states' AI exposure (see SI Section 2) and states' job separation or total unemployment rate (see Fig.~\ref{fig:individual}A\&B, respectively).
On aggregate, Webb's~\cite{webb2019impact} measures for Robotics exposure were most predictive of job separation rates while measures from Acemoglu \& Autor~\cite{acemoglu2011skills} were most predictive of unemployment rates (see Fig.~\ref{fig:individual}D).
But each individual score contributed only a small amount of variance explained compared to the baseline model.
However, combining all scores into a single model yields a 16.8\% improvement in explained total unemployment rate variation by state and month and a 20.2\% improvement for job separation rates by state and month according to out-of-sample cross-validation (see Fig.~\ref{fig:crossValidation}).
Full regression tables are provided in SI Sections 6 and 7.

Does technology shape labor outcomes by changing skill demands?
Technology rarely automates entire occupations wholesale but instead automates specific workplace activities~\cite{frank2019toward,acemoglu2011skills}.
These changes may be subtle without producing job separations or unemployment if workers adapt.
Some studies hypothesize that AI~\cite{webb2019impact,acemoglu2021harms} or machine learning (ML)~\cite{brynjolfsson2018what} would change skill demands within occupations slowly enough for workers to adapt their skills and/or find new employment.
However, re-skilling or up-skilling still imposes an adjustment cost on both workers and employers~\cite{brynjolfsson2021productivity,tambe2020digital}.

We use occupation skill profiles from the BLS O*NET database to track changes to skill demands within about 700 different occupations (see SI Section 8).
Routine Manual work was the only Wave 1 AI exposure score positively associated with skill change (see Fig.~\ref{fig:individual}C).
From Wave 2, both Frey \& Osborne's~\cite{frey2017the} and Arntz et al.'s~\cite{arntz2016risk} scores were positively associated with skill change.
Technology-specific exposure scores from Wave 3 yielded mixed results.
Suitability for ML~\cite{brynjolfsson2018what} predicts greater skill change, while AI exposure from Felten et al.~\cite{felten2018method} predicts less change.
This is especially surprising because these studies employed similar methodologies; both studies surveyed Mechanical Turk users about the specific capabilities of AI or ML to produce their occupation scores.
Ultimately, AI exposure according to Felten et al.~\cite{felten2018method} yielded the largest gain in predictive performance, adding 5.7\% variance explained (i.e., adjusted $R^2$) compared to the baseline fixed effects model (see Fig.~\ref{fig:individual}D).
However, adding the combined AI exposure model to a baseline model controlling for year and occupation fixed effects produces a 58.8\% improvement in variation explained (see Fig.~\ref{fig:crossValidation}).
Summary skill change statistics and full regression tables are provided in SI Section 8.

\section*{Where and When Automation Estimates Work}

\begin{figure}[t]
    \centering
    \includegraphics[width=\textwidth]{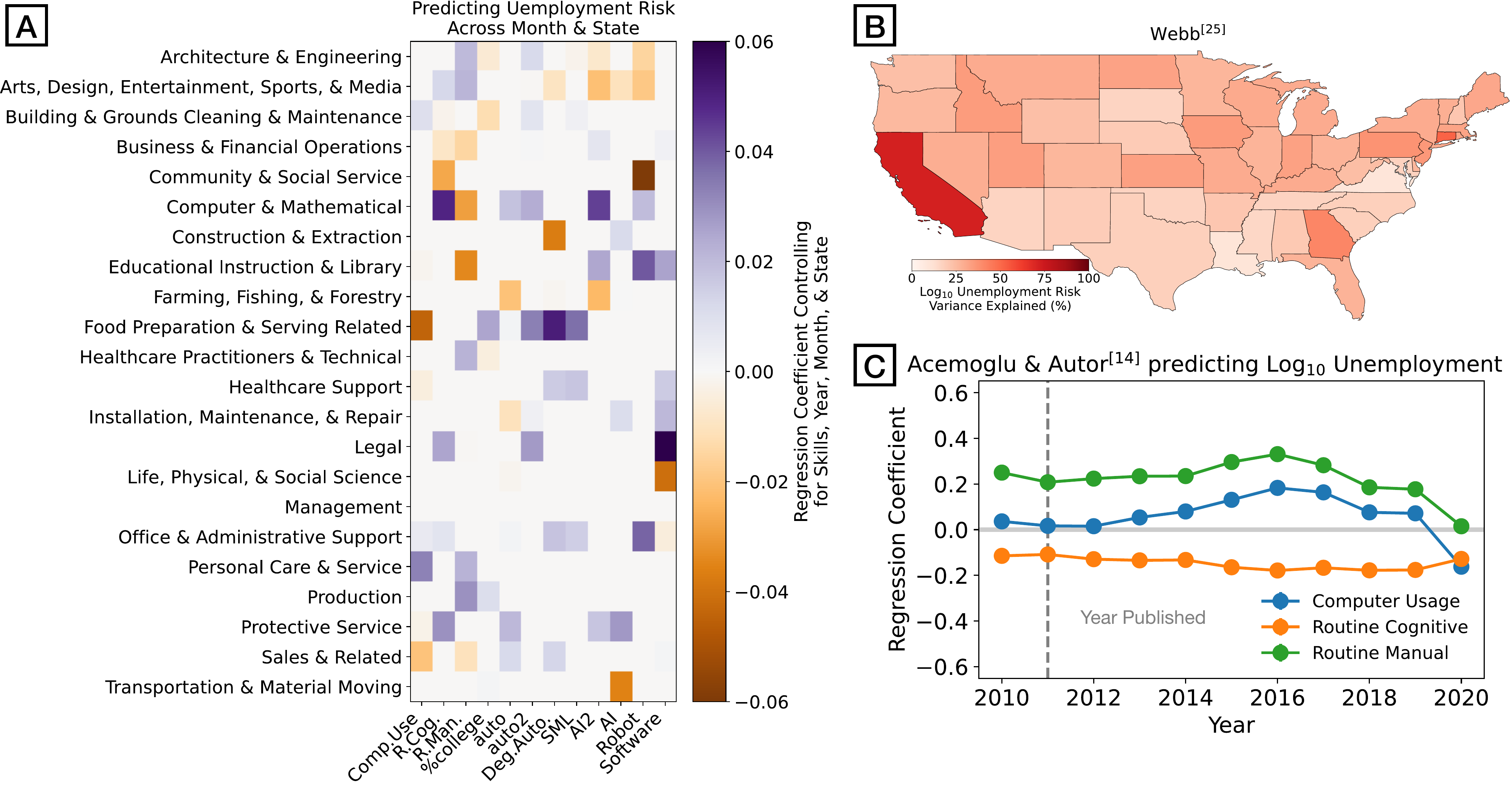}
    \caption{
        The effectiveness of AI exposure scores varies by occupation, year, and state.
        {\bf (A)} AI exposure scores' correspondence with unemployment risk across states and months.
        Individual regressions were run for each pair of major occupation and AI exposure scores while controlling for skills, month, year, and state.
        We report regression coefficient estimates with $p$-value $<10^{-2}$.
        {\bf (B)} The performance of scores from Webb~\cite{webb2019impact} in combination with year and month fixed effects for predicting unemployment risk by occupation.
        Individual regressions were performed for each state.
        {\bf (C)} For each year, we regress scores from Acemoglu \& Autor~\cite{acemoglu2011skills} against unemployment risk by occupation with state and month-fixed effects. 
        We provide 95\% confidence intervals for coefficient estimates.
        In all plots, all variables were centered and standardized before analysis.
        Similar plots for other AI exposure scores are provided in the Supplementary Materials.
    }
    \label{fig:temporal}
\end{figure}

Examining each AI exposure score's performance across occupations, locations, and time periods reveals how each score contributes to the combined model.
For example, unemployment risk for Sales workers across states was negatively associated with the Routine Cognitive score~\cite{acemoglu2011skills} and positively associated with Suitability for ML~\cite{brynjolfsson2018what} (see Fig.~\ref{fig:temporal}A. Analysis included year and month fixed effects).
As an example of heterogeneous spatial performance, combined with year and month fixed effects, \% AI, Robotics, and Software scores from Webb~\cite{webb2019impact} are most predictive of unemployment risk by occupation in California compared to other US states (see Fig.~\ref{fig:temporal}B).
However, other AI exposure scores yielded more even performance across states.
For example, state-level scores derived from Frey \& Osborne~\cite{frey2017the} yielded roughly equal predictive performance across states when predicting total job separation rates (see SI Section 7 for similar maps of other exposure scores).
Similarly, the performance of individual exposure scores can vary over time.
For example, Computer Usage~\cite{acemoglu2011skills} was increasingly predictive of unemployment risk before 2016 but became less strongly associated with risk thereafter---even becoming negatively associated with unemployment risk in 2020 perhaps due to more work-from-home activities during the COVID pandemic (see Fig.~\ref{fig:temporal}C).
As another example, using scores from Webb~\cite{webb2019impact}, percentage exposure to Robots indicated greater total job separation rates in states except for the year 2017 when exposure to Software became positively associated with separations (see SI Section 7).
Thus, the best model for predicting labor outcomes may vary depending on the spatial or industrial context.

\section*{Discussion}

How does technology shape labor outcomes?
Previous AI exposure studies only used occupations' employment or wage trends to evaluate exposure measures.
However, other labor data should be considered if policy makers, workers, and employers want to prepare for the future of work.
This study validates AI exposure scores against many more possible labor outcomes besides just employment or wage changes.
For instance, AI exposure studies~\cite{frank2019toward,acemoglu2011skills,webb2019impact,acemoglu2021harms,brynjolfsson2018what,mean2023job} claim that within-occupation skill change is the most likely outcome.
Our study supports this claim by showing that AI exposure scores yields a 59\% factor improvement in skill change predictions (see Fig.~\ref{fig:crossValidation}).
However, in disagreement with claims that technology does not impact unemployment~\cite{autor2015there}, our results also associate AI exposure with increased job separation rates in states and unemployment risk across occupations. 
Thus, employment and wage data are not enough to fully quantify the labor implications of technological change.
Future research on technology and automation can pursue a more comprehensive study of technology and labor using this study's data on unemployment risk by occupation, state, and month (see \url{https://github.com/mrfrank8176/unemploymentRisk}).

As federal and state policies aim to prepare for new technology and the future of work (e.g., Illinois Future of Work Act 2021, 2021 IL H 645, and the National Artificial Intelligence Initiative Act of 2020 (NAIIA)), how should government and policy analysts quantify AI exposure?
More than a decade of research has produced multiple efforts estimating occupations' AI exposure and these exposure scores have been used in government and industry analyses, including by the Brookings Institution~\cite{muro2017digitalization}, the US Bureau of Labor Statistics~\cite{mean2023job}, and the Organisation for Economic Co-operation and Development (OECD)~\cite{nedelkoska2018automation,georgieff2021happened,lane2021impact,georgieff2021artificial}. 
The lack of agreement among different scores and their poor individual ability to predict unemployment risk, job separation rates, or within-occupation skill change (see Fig.~\ref{fig:individual}) suggest that more research is required to converge on good solutions with practical use in policy research.

How can we improve our understanding of work disruptions from technology?
More recent scores from Wave 3 were more predictive of occupations' unemployment risk (see Fig.~\ref{fig:scores}), suggesting that focusing on technologies' specific capabilities may yield further improvements.
But, on aggregate, the significant predictive performance of the combined model (see Fig.~\ref{fig:crossValidation}) demonstrates that existing scores identify key factors when taken together.
A better understanding of why AI exposure scores perform differently depending on the state, year, and occupation (see Fig.~\ref{fig:temporal}) could yield progress.
For example, the high performance of scores from Webb's study~\cite{webb2019impact} in California may be expected because of the study's focus on information technology and the technology industry's presence in Silicon Valley.
Similarly, the scores used in this study were published across the past decade, and their association with labor outcomes across states and/or occupations varies over time.
That is, we use scores from existing studies in the last decade to predict labor outcomes throughout the entire time period regardless of the study's publication year (i.e., all scores are used for both prospective and retrospective analysis).
Combined, these results demonstrate a need for more research and development in measuring AI exposure.
Future work might produce dynamic AI exposure scores that update annually based on occupations' skill requirements, new technological capabilities (e.g., according to recent patents), or based on individual outcomes~\cite{fossen2022new}.
More research may identify composite AI exposure scores that leverage the methodological strengths of studies across multiple waves.

Our study has limitations.
First, our analysis of unemployment risk used characteristic unemployment data for occupation groups (i.e., Standard Occupation Classification Major Occupations) when more granular occupation data would be preferred.
Similarly, our analysis of job separation rates was limited to the aggregation of states rather than by occupation, industry, or firm.
Despite these shortcomings, to the best of our knowledge, there are no better comprehensive, nationally-representative data for describing unemployment risk or job separations using federal data.
Refining data from the Department of Labor~\cite{dell2020assessing,mean2023job} will spur dramatic improvements in predictions of technology and the future of work.
Second, similar to previous AI exposure studies, this study does not establish causality between AI exposure and labor outcomes because it is extremely difficult to identify natural experiments around AI exposure at the national scale.
Endogeneity exists because most nascent technologies become mature enough to impact the national labor system only if they fill some existing demand in the labor market.
That is, new technologies that are in demand will receive enough resources to evolve from research and development to production and deployment.
Although our study fails to control for every potential confound, our analyses do control for year (i.e., long-term trends), month (i.e., seasonality), and state (i.e., geographical) fixed effects where appropriate.
For example, unlike previous efforts to validate AI exposure scores, our analysis of unemployment risk by occupation, state, and month controls for occupations' skill requirements and education requirements according to the US BLS (see Fig.~\ref{fig:combined}A).
Many factors, such as workers' self-selection into different types of work or increased global competition~\cite{autor2013china,walter2017globalization}, will confound the relationship between AI exposure and unemployment risk through skills or education requirements and, thus, would be at least partially controlled for in our analysis.
Third, there are many other potential outcomes to AI exposure including changes to returns on education~\cite{huseynov2023chatgpt} and retirement decisions~\cite{casas2023early}.
Fourth, our study is focused on \emph{digital} AI and misses AI applications in other domains, including robotics~\cite{paolillo2022compete,acemoglu2020robots}.
Fifth, our study does not account for recent AI applications whose impact on the labor market are yet to materialize.
However, even recent and future AI applications and associated efforts to measure workers' exposure~\cite{eloundou2023gpts,felten2023occupational} require careful validation against detailed labor data.

In conclusion, we have demonstrated that many concerning labor outcomes can be predicted with an ensemble model of AI exposure scores.
Examples include unemployment risk, skill change per occupation, and job separation rates by state.
While policymakers strive to prepare workers for artificial intelligence and the future of work, new methods are required to holistically quantify which workers have exposure to technology and the risk of detrimental labor outcomes.

\section*{Acknowledgements}
This research is partly supported by the University of Pittsburgh Momentum Fund and the Center for Research Computing. 
E.M. is in part supported by the National Science Foundation under Grant No.~2218748, and by Ministerio de Ciencia e Innovaci\'on/Agencia Espa\~nola de Investigaci\'on (MCIN/AEI/10.13039/501100011033), Spain through grant PID2019-106811GB-C32. 
Y.Y.A. is in part supported by the National Science Foundation Grant No.~2241237. 
We thank Byungkyu Lee, Daniel Rock, Jaehyuk Park, Maria del Rio-Chanona, and Dakota Murray for their helpful comments on the manuscript.

\bibliographystyle{unsrt}
\bibliography{main}

\newpage

\section{Supplementary Materials: AI exposure predicts unemployment risk}

\subsection{Estimating Technology Exposure}
\label{SI:history}
Estimations of technology exposure have evolved over the last decade.
The first wave of theoretical studies adapted a production function~\cite{autor2008trends,acemoglu2002technical} to worker productivity in the presence of automating technologies, arguing that college-educated cognitive high-skill workers were complemented by technology, including computers, while manual low-skill workers were substituted by technologies like robotics.
However, current AI technologies threaten cognitive workers as well. 
As examples, consider that AI surpasses human performance at predicting heart attacks~\cite{hutson2017self} or computer vision applications in radiology~\cite{ranschaert2019advantages,king2018artificial,schier2018artificial} and neighborhood safety~\cite{naik2017computer}.
The modern, skill-biased technological change framework~\cite{acemoglu2011skills} further argues that both routine manual and cognitive work are ripe for automation---although cognitive workers will tend towards greater productivity with technology while manual workers will tend towards labor substitution (e.g., in manufacturing~\cite{acemoglu2020robots}).

The second wave of studies considered each occupation as a bundle of skill requirements and job tasks.
Beyond describing occupations as cognitive or routine, occupations' granular skill requirements are considered, for instance, by leveraging the US Bureau of Labor Statistics (BLS) O*NET database~\cite{frank2019toward,alabdulkareem2018unpacking,handel2016net}.
An Oxford University study~\cite{frey2017the} subjectively identified ``fully automatable'' and ``not automatable'' occupations combined with a subset of O*NET variables representing perception, manipulation, creativity, and social intelligence requirements of occupations.
They employed a logistic regression to assign a ``probability of computerisation'' to each remaining US occupation.
Alarmingly, they claimed that 47\% of US employment had high risk of computerization.
However, the study only compared their estimates to occupations' education requirements and wages, thus leaving whether this exposure creates unemployment or alters skill demands unclear.
A competing study directly estimated the automation risk of skills from the Programme for the International Assessment of Adult Competencies (PIAAC) survey enabling a direct assessment of technological exposure of occupations across OECD countries~\cite{arntz2016risk}, finding that only 9\% of US workers had high automation risk.
Subsequent studies used these occupation estimates in a variety of contexts; for example, finding that automation will affect 35\% of employment in Finland~\cite{pajarinencomputerization2015}, 59\% of employment in
Germany~\cite{brzeski2015roboter}, 45 to 60\% of employment across Europe~\cite{bowles2014computerisation}, and that small US cities face greater impact from automation~\cite{frank2018small}.
To meet the demand for occupation-level automation estimates, the BLS added a Degree of Automation score to occupation profiles in the 2016 O*NET database.

The most-recent third wave of studies directly connects specific technological capabilities to occupations' job tasks to assess each occupation's exposure (i.e., a task-based approach~\cite{acemoglu2021tasks}).
One study surveyed machine learning (ML) experts on the characteristics of tasks that are suitable for ML and produced a Suitability for ML score for US occupations~\cite{brynjolfsson2018what}.
Another study surveyed gig workers to establish connections between AI application capabilities and occupations~\cite{felten2018method}.
A more recent study used natural language processing to connect technology patents to job tasks~\cite{webb2019impact}.
Although motivated by the risk of technological unemployment, these studies argue that AI exposure will mostly result in labor reorganization through wealth inequality or changing skill demands without major changes to unemployment.

\subsection{Technology Exposure by State}
\label{SI:stateExposure}
We study monthly job separation and total unemployment rates by state using the Job Openings and Labor Turnover Survey (JOLTS) and 
Local Area Unemployment Statistics (LAUS) from the US Bureau of Labor Statistics (BLS).
Unfortunately, these data are not stratified by state and occupation, but they are representative of each state's monthly economy.
Given a per-occupation technology exposure score $exposure(j)$, we calculate the aggregate exposure for state $s$ in year $y$ according to 
\begin{equation}
    exposure(s,y) = \sum_{j\in SOC} exposure(j)\cdot share_{s,y}(j)
    \label{eq:stateScores}
\end{equation}
where $j$ is a six-digit Standard Occupation Classification code and $share_{s,y}(j)$ is the share of employment associated with occupation $j$ in that state and year according to the BLS.

\subsection{Quantifying Occupations' Unemployment Risk and States' Job Separations}
\label{SI:data}
This study uses federal government data spanning 2010 through 2020 (except where a different time frame is noted).
The US Bureau of Labor Statistics (BLS) Local Area Unemployment Statistics (LAUS) provides total unemployment rates, and the BLS Job Openings and Labor Turnover Survey (JOLTS) provides total job separation rates (i.e., fires and quits) by state and month.
These data capture the aggregate labor dynamics in each state but obfuscate the heterogeneous dynamics of workers across occupations.
Evaluating technology exposure additionally requires high-resolution per-occupation data about unemployment risk and changes to skill demands.
We meet this requirement with data describing occupations according to the Standard Occupation Classification (SOC) system.
The SOC is used by the US Department of Labor as a standardized taxonomy of occupations.
Each of about 700 occupations is specified by a unique six-digit code where the first two digits also indicate the occupation's \emph{major} occupation group (hereafter, ``major occupation'').
For example, the BLS Occupational Employment and Wage Statistics (OEWS) details the employment distribution over six-digit occupations in each state each year.
The BLS also provides the O*NET database, which provides annual profiles of the skills, knowledge, abilities, and work contexts required by each six-digit SOC code.

\begin{figure}[H]
    \centering
    \includegraphics[width=.5\textwidth]{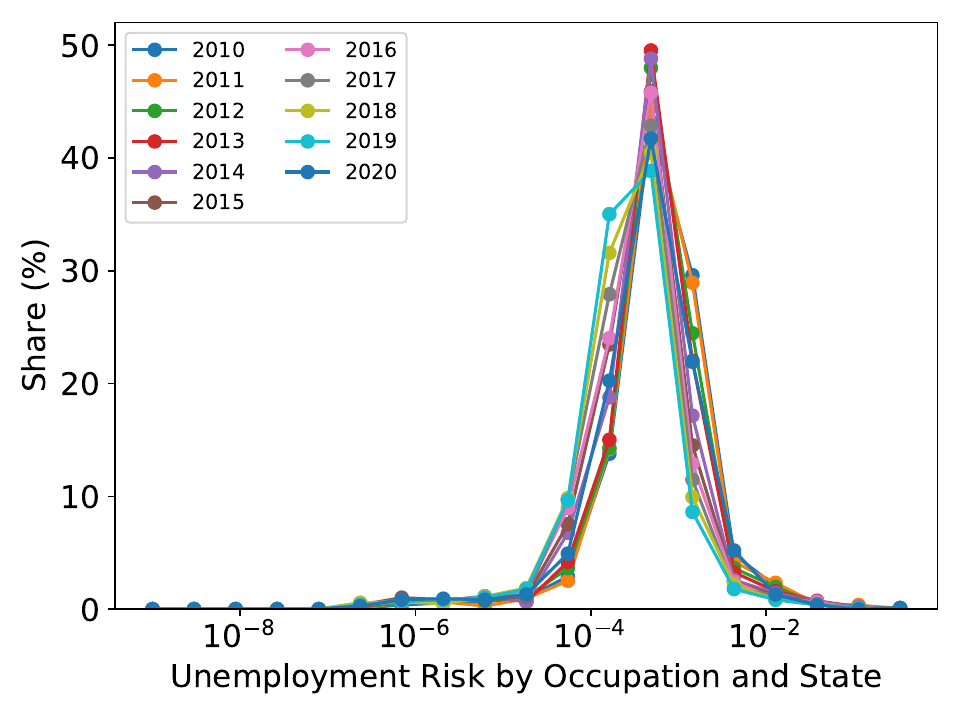}
    \caption{
        Annual distributions of occupation unemployment risk.
        The median of monthly risk scores by Major SOC code and state within each year (i.e., median taken over monthly values within each year).
    }
    \label{SI:riskDist}
\end{figure}

These data include occupation employment and wages but lack insights into unemployment by occupation.
We overcome this barrier using data from the US Department of Education and Training Administration (DETA).
The DETA reports characteristic unemployment data from each US state's unemployment insurance department detailing the most recent major occupation of every benefit recipient in every state every month.
Using this data, given the state $s$ and month $m$, we directly compute the probability that an unemployment recipient's most recent major occupation was $soc$ (denoted $p_{s,m}(soc|u)$).
In combination with annual state employment distributions from OEWS, we use Bayes' Theorem to estimate each major occupation's \emph{unemployment risk} in state $s$ and month $m$ according to 
\begin{equation}
    p_{s,m}(u|soc) = p_{s,m}(soc|u)\cdot p_{s,m}(u)/p_{s,m}(soc).
    \label{eq:uiRisk}
\end{equation}
$p_{s,m}(u)$ is the aggregate unemployment rate in each state according to LAUS.
$p_{s,m}(soc)$ is estimated according to the probability that a worker in a state's labor force is either actively employed in $soc$ according to annual OEWS data or is unemployed and was most recently employed in $soc$ according to monthly DETA statistics. 
Note that $p_{s,m}(soc)$ is relative to the state's labor force rather than just employed workers.
We provide distributions of unemployment risk estimates in Figure~\ref{SI:riskDist}.

\subsection{O*NET Principal Component Analysis}
\label{SI:onetPcaSection}
\begin{figure}[H]
    \centering
    \includegraphics[width=.5\textwidth]{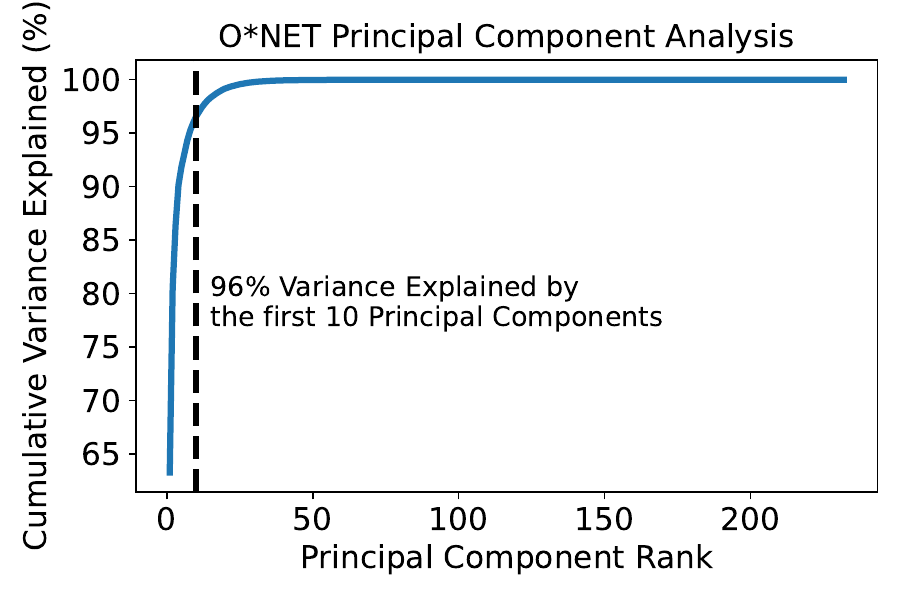}
    \caption{
        O*NET variance explained by the number of principal components included.
        In this study, we use the first ten principal components which cumulatively account for 96\% of the overall variation in occupations' skill requirements by year according to O*NET.
    }
    \label{SI:onetPcaVarExp}
\end{figure}

We quantify occupation's skill requirements using the US Bureau of Labor Statistic's (BLS) O*NET skills database.
These occupation skill profiles result from nationally-representative worker surveys covering over 700 different job titles from the Standard Occupation Classification (SOC) system and details the importance of 120 to 230 examples of workplace knowledge, abilities, work activities, and skills (henceforth, ``skills'') in each year from 2010 through 2020.
Each occupation is identified by a unique six-digit SOC code where the first two-digits describe the occupation's Major Occupation type.
We use $onet_{y,j,s}\in[0,1]$ to denote the importance of skill $s\in S$ to occupation $j\in J$ in year $y$ such that $onet_{y,j,s}=1$ identifies an essential skill and $onet_{y,j,s}=0$ indicates an irrelevant skill.
We simplify these annual occupation skill profiles using principal component analysis (PCA).
Figure~\ref{SI:onetPcaVarExp} demonstrates the variation in skill requirements captured by PCA as a function of the number of principal components included in the analysis.
Throughout this study, we use the first ten principal components to control for occupations' skill requirements.
The first ten principal components account for 96\% of the total variation in occupation's skill requirements across all years of the O*NET data.
These principal components are used in our analysis of unemployment risk to control for occupations' skill requirements.

\subsection{Unemployment Risk by Occupation, State, \& Month}
\label{SI:risk}

\begin{figure}[H]
    \centering
    \includegraphics[width=.8\textwidth]{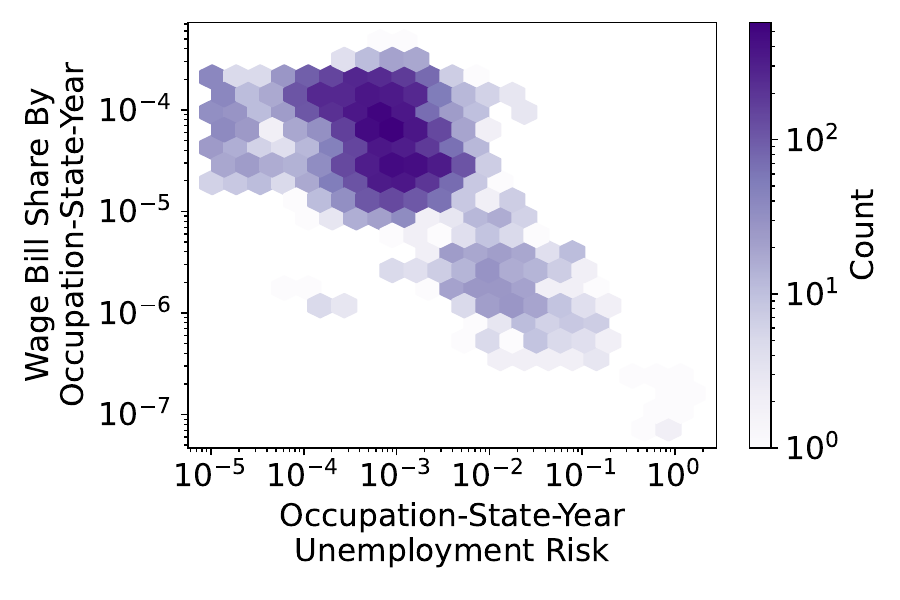}
    \caption{
        Using occupation wage data, we observe a negative correlation between unemployment risk and wages.
        BLS OEWS describes occupation wages for each six-digit SOC code, state, and year.
        However, unemployment risk is only calculated for each two-digit SOC code (i.e., Major Occupation), state, and year.
        We sum six-digit SOC wage bill values by state to calculate wage bill values for two-digit SOC codes in each state in each year.
    }
    \label{SI:wagesVsUi}
\end{figure}

\begin{table}[H]
    \centering
    \fontsize{6.5}{6}\selectfont
\begin{tabular}{l|llllllll}
    \hline
    \multicolumn{9}{c}{Dependent Variable: Log$_{10}$ Unemployment Risk by Occupation, Month, \& State }\\ \hline
    Variable &        Model 1 &        Model 2 &       Model 3 &       Model 4 &       Model 5 &       Model 6 &        Model 7 &        Model 8 \\
         \hline Acemoglu \& Autor$^{[24]}$ Comp.Use &          0.000 &                &               &               &               &               &                &                \\
                  Acemoglu \& Autor$^{[24]}$ R.Cog. & -0.096$^{***}$ &                &               &               &               &               &                &                \\
                  Acemoglu \& Autor$^{[24]}$ R.Man. &  0.137$^{***}$ &                &               &               &               &               &                &                \\
                           ONET Education \%college &                & -0.134$^{***}$ &               &               &               &               &                &                \\
                       Frey \& Osborne$^{[9]}$ auto &                &                & 0.024$^{***}$ &               &               &               &                &                \\
                        Arntz et al$^{[10]}$ auto2 &                &                &               & 0.327$^{***}$ &               &               &                &                \\
                         ONET Automation Deg.Auto. &                &                &               &               & 0.152$^{***}$ &               &                &                \\
                   Brynjolfsson et al$^{[11]}$ SML &                &                &               &               &               & 0.082$^{***}$ &                &                \\
                         Felten et al$^{[12]}$ AI2 &                &                &               &               &               &               & -0.109$^{***}$ &                \\
                                  Webb$^{[13]}$ AI &                &                &               &               &               &               &                &  0.332$^{***}$ \\
                               Webb$^{[13]}$ Robot &                &                &               &               &               &               &                &  0.412$^{***}$ \\
                            Webb$^{[13]}$ Software &                &                &               &               &               &               &                & -0.294$^{***}$ \\
                                             \hline
$R^2$ &          0.028 &          0.018 &         0.001 &         0.107 &         0.023 &         0.007 &          0.012 &          0.089 \\
                                        adj. $R^2$ &          0.028 &          0.018 &         0.001 &         0.107 &         0.023 &         0.007 &          0.012 &          0.089 \\
\hline \multicolumn{9}{c}{$p_{val}<0.1^*$, $p_{val}<0.01^{**}$, $p_{val}<0.001^{***}$} \\ \hline
\end{tabular}
    \caption{
        Linear regression analysis of occupations' technology exposure and unemployment risk.
        Data varies by Major SOC code, state, and month from January 2010 through 2020 for a total of 140,274 data points.
        All variables were centered and standardized before analysis, thus eliminating units from the various exposure variables.
    }
    \label{SI:riskRegressionSimple}
\end{table}

\begin{table}[H]
    \centering
    \fontsize{7}{7}\selectfont
\begin{tabular}{l|lllllllll}
\hline\multicolumn{10}{c}{Dependent Variable: Log$_{10}$ Unemployment Risk by Occupation, Month, \& State }\\ \hline
                                              Variable &       Model 1 &        Model 2 &        Model 3 &       Model 4 &       Model 5 &        Model 6 &        Model 7 &        Model 8 &        Model 9 \\
         \hline Acemoglu \& Autor$^{[24]}$ Comp.Use &               &  0.715$^{***}$ &                &               &               &                &                &                &  0.534$^{***}$ \\
                  Acemoglu \& Autor$^{[24]}$ R.Cog. &               & -0.510$^{***}$ &                &               &               &                &                &                & -0.241$^{***}$ \\
                  Acemoglu \& Autor$^{[24]}$ R.Man. &               &          0.000 &                &               &               &                &                &                &          0.000 \\
                       Frey \& Osborne$^{[9]}$ auto &               &                & -0.582$^{***}$ &               &               &                &                &                & -0.317$^{***}$ \\
                        Arntz et al$^{[10]}$ auto2 &               &                &                & 0.245$^{***}$ &               &                &                &                &  0.377$^{***}$ \\
                         ONET Automation Deg.Auto. &               &                &                &               & 0.054$^{***}$ &                &                &                &          0.000 \\
                   Brynjolfsson et al$^{[11]}$ SML &               &                &                &               &               & -0.102$^{***}$ &                &                & -0.198$^{***}$ \\
                         Felten et al$^{[12]}$ AI2 &               &                &                &               &               &                & -0.165$^{***}$ &                &  0.014$^{***}$ \\
                                  Webb$^{[13]}$ AI &               &                &                &               &               &                &                &  0.468$^{***}$ &  0.114$^{***}$ \\
                               Webb$^{[13]}$ Robot &               &                &                &               &               &                &                &  0.291$^{***}$ &  0.647$^{***}$ \\
                            Webb$^{[13]}$ Software &               &                &                &               &               &                &                &         -0.000 & -0.098$^{***}$ \\
                           \hline ONET Education \%college & 0.032$^{***}$ & -0.124$^{***}$ & -0.032$^{***}$ & 0.061$^{***}$ &         0.000 &  0.075$^{***}$ &  0.038$^{***}$ & -0.116$^{***}$ & -0.036$^{***}$ \\
                                         O*NET PCA &           Yes &            Yes &            Yes &           Yes &           Yes &            Yes &            Yes &            Yes &            Yes \\
                                         Year F.E. &           Yes &            Yes &            Yes &           Yes &           Yes &            Yes &            Yes &            Yes &            Yes \\
                                        Month F.E. &           Yes &            Yes &            Yes &           Yes &           Yes &            Yes &            Yes &            Yes &            Yes \\
                                        State F.E. &           Yes &            Yes &            Yes &           Yes &           Yes &            Yes &            Yes &            Yes &            Yes \\
                                             \hline
$R^2$ &         0.579 &          0.690 &          0.690 &         0.587 &         0.580 &          0.586 &          0.593 &          0.657 &          0.762 \\
                                        adj. $R^2$ &         0.579 &          0.689 &          0.690 &         0.587 &         0.580 &          0.586 &          0.592 &          0.657 &          0.762 \\
\hline \multicolumn{10}{c}{$p_{val}<0.1^*$, $p_{val}<0.01^{**}$, $p_{val}<0.001^{***}$} \\ \hline
\end{tabular}
    \caption{
        LASSO regression analysis of occupations' technology exposure and unemployment risk.
        Data varies by Major SOC code, state, and month from January 2010 through 2020 for a total of 140,274 data points.
        All variables were centered and standardized before analysis, thus eliminating units from the various AI exposure variables.
    }
    \label{SI:riskRegression}
\end{table}

\newcounter{ct}
\begin{figure}[H]
    \centering
    \forloop{ct}{0}{\value{ct} < 8}{
        \includegraphics[width=.4\textwidth]{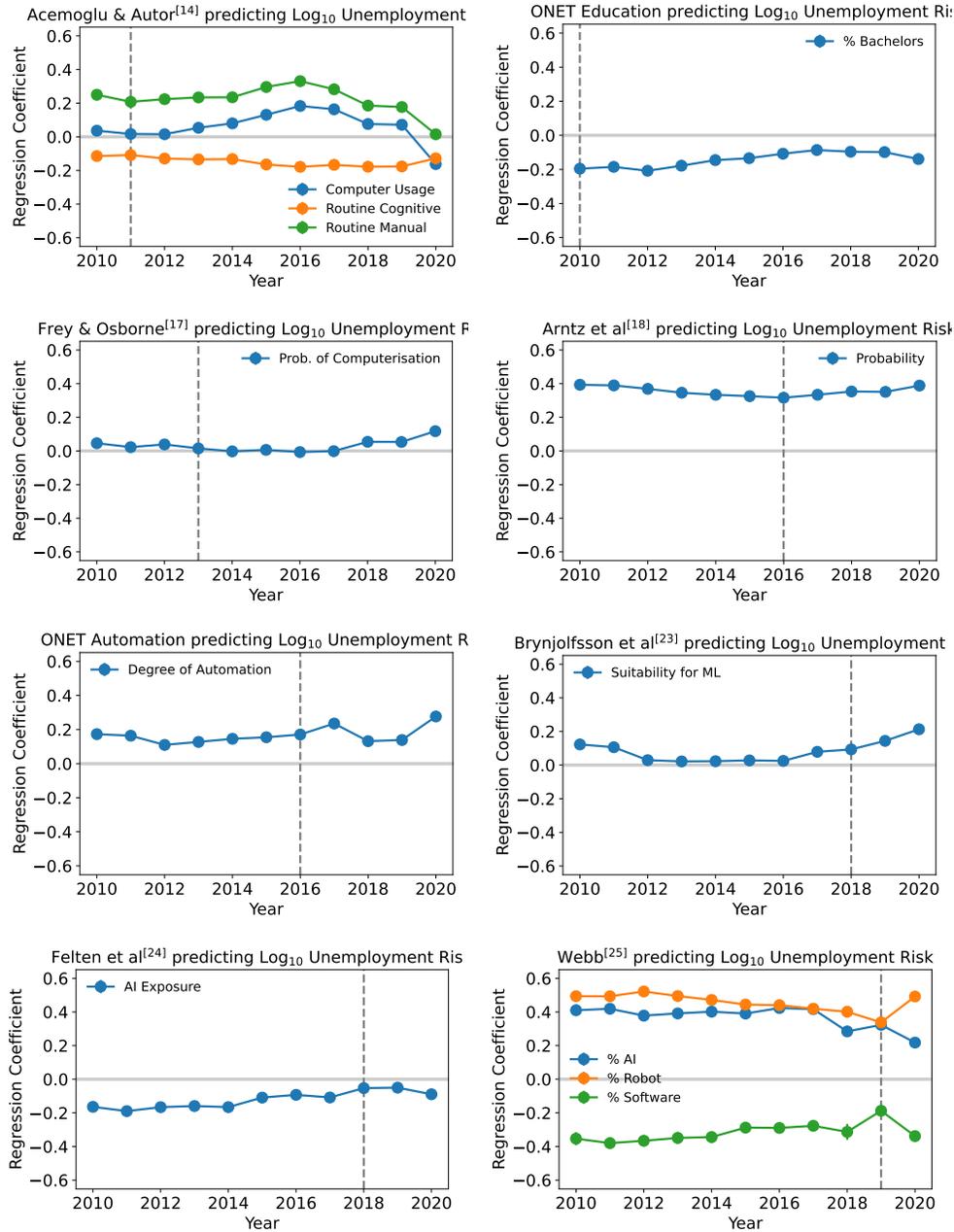}
    }
    \caption{
        Variables' regression coefficient estimate with $\log_{10}$ unemployment risk as the dependent variable.
        Bars represent 95\% confidence intervals.
        Independent regressions with month and state fixed effects were performed for each year.
        All variables were centered and standardized before analysis, thus eliminating units from the various AI exposure variables.
    }
    \label{SI:uiCoefOverTime}
\end{figure}

\begin{figure}[H]
    \centering
    \forloop{ct}{0}{\value{ct} < 8}{
        \includegraphics[width=.4\textwidth]{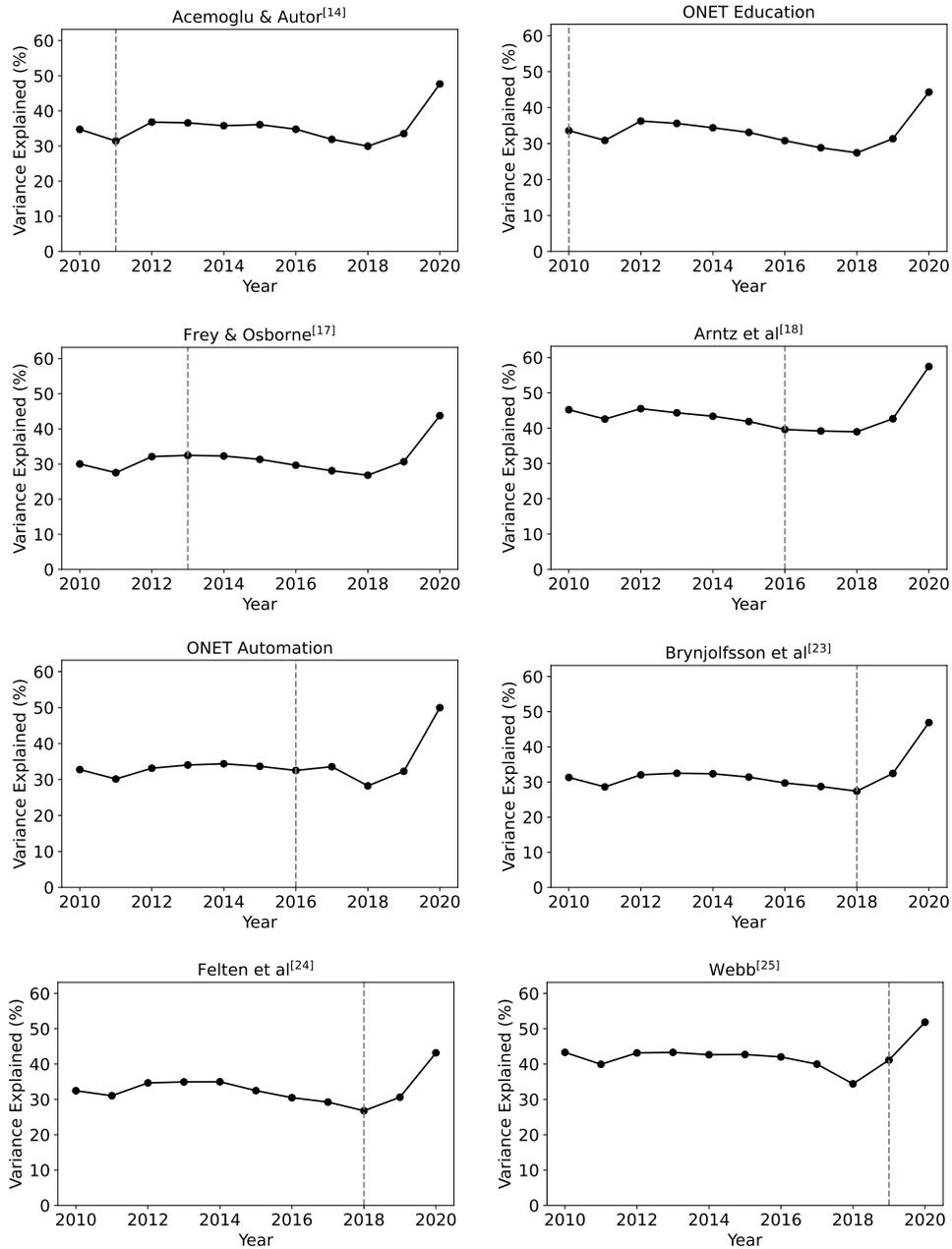}
    }
    \caption{
        $\log_{10}$ unemployment risk variance explained by each model.
        Independent regressions with month and state fixed effects were performed for each year.
        All variables were centered and standardized before analysis, thus eliminating units from the various AI exposure variables.
        Vertical dashed line represents the year that AI exposure variables were made available.
    }
    \label{SI:uiROverTime}
\end{figure}

\begin{figure}[H]
    \centering
    \forloop{ct}{0}{\value{ct} < 8}{
        \includegraphics[width=.49\textwidth]{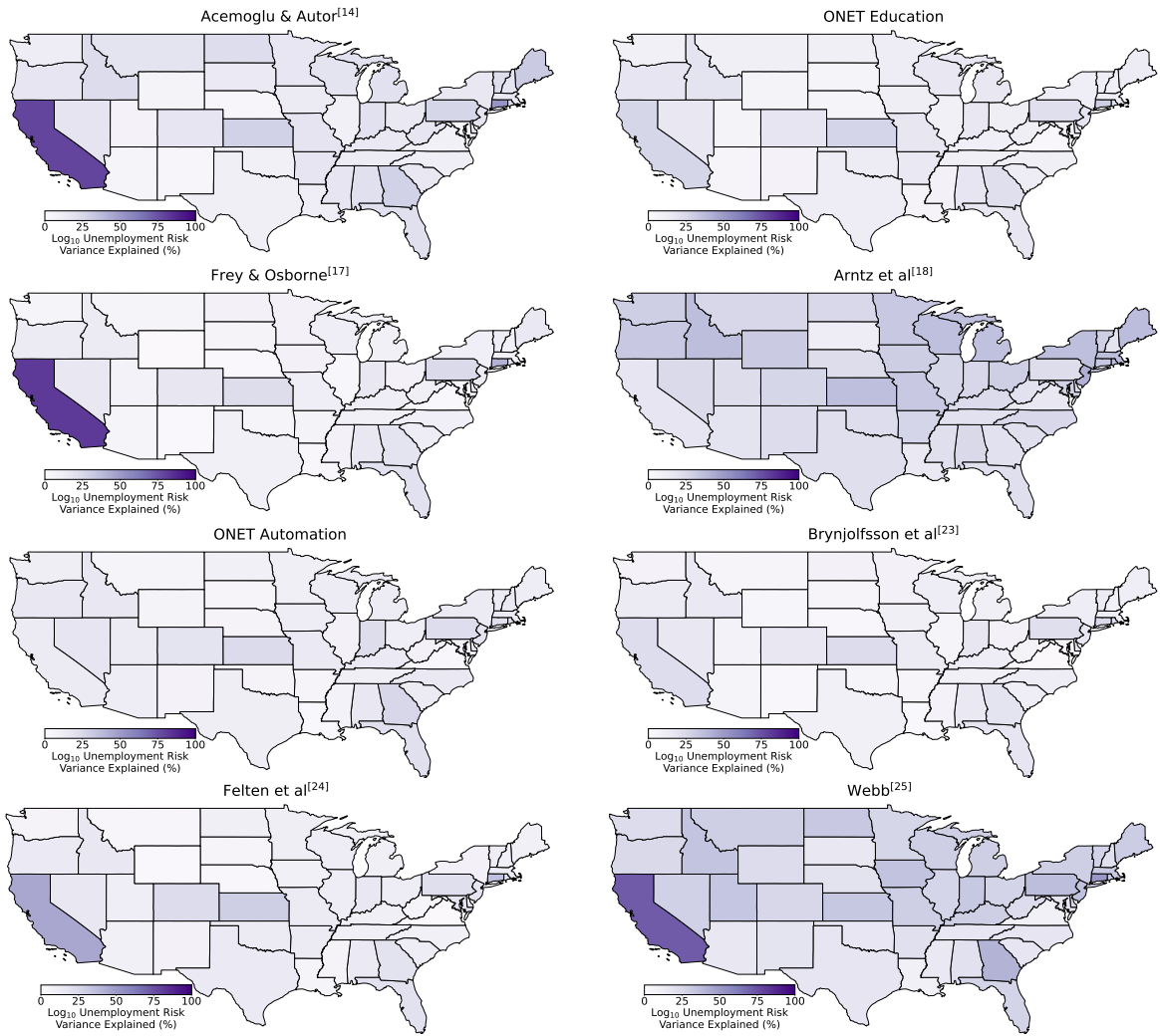}
    }
    \caption{
        $\log_{10}$ unemployment risk variance explained by each model.
        Independent regressions with month and year fixed effects were performed for each state.
    }
    \label{SI:uiRMap}
\end{figure}

\subsection{State Unemployment Rates}
\label{SI:unemployment}
\begin{table}[H]
    \centering
    \fontsize{6.5}{7}\selectfont
    \begin{tabular}{l|ll|lll|lll}
    \hline
    \multicolumn{9}{c}{Dependent Variable: Log$_{10}$ Total Separations by State \& Month} \\ \hline
        & \multicolumn{2}{c|}{Wave 1} & \multicolumn{3}{c|}{Wave 2} & \multicolumn{3}{c}{Wave 3} \\ 
                                                          Variable &        Model 1 &        Model 2 &        Model 3 &        Model 4 &        Model 5 &        Model 6 &        Model 7 &        Model 8 \\ \hline
          Acemoglu \& Autor$^{[14]}$ Computer Usage &  0.032$^{*  }$ &                &               &               &               &               &               &               \\
       Acemoglu \& Autor$^{[14]}$ Routine Cognitive &  0.458$^{***}$ &                &               &               &               &               &               &               \\
          Acemoglu \& Autor$^{[14]}$ Routine Manual & -0.323$^{***}$ &                &               &               &               &               &               &               \\
             ONET Education \% Workers w/ Bachelors &                & -0.073$^{***}$ &               &               &               &               &               &               \\
  Frey \& Osborne$^{[17]}$ Prob. of Computerisation &                &                & 0.247$^{***}$ &               &               &               &               &               \\
                  Arntz et al$^{[18]}$ Probability &                &                &               & 0.168$^{***}$ &               &               &               &               \\
              ONET Automation Degree of Automation &                &                &               &               & 0.312$^{***}$ &               &               &               \\
    Brynjolfsson et al$^{[23]}$ Suitability for ML &                &                &               &               &               & 0.185$^{***}$ &               &               \\
                 Felten et al$^{[24]}$ AI Exposure &                &                &               &               &               &               & 0.264$^{***}$ &               \\
                                Webb$^{[25]}$ \% AI &                &                &               &               &               &               &               &        -0.013 \\
                             Webb$^{[25]}$ \% Robot &                &                &               &               &               &               &               &         0.046 \\
                          Webb$^{[25]}$ \% Software &                &                &               &               &               &               &               & 0.101$^{*  }$ \\
                                             \hline
$R^2$ &          0.176 &          0.005 &         0.061 &         0.028 &         0.097 &         0.034 &         0.069 &         0.018 \\
                                        adj. $R^2$ &          0.175 &          0.005 &         0.061 &         0.028 &         0.097 &         0.034 &         0.069 &         0.018 \\
\hline \multicolumn{9}{c}{$p_{val}<0.1^*$, $p_{val}<0.01^{**}$, $p_{val}<0.001^{***}$} \\ \hline
\end{tabular}
    \caption{
        Linear regression analysis of states' technology exposure and total unemployment rates according to LAUS.
        Data varies by state, and month from January 2010 through 2020 for a total of 6,600 data points.
        All variables were centered and standardized before analysis thus eliminating units from the various exposure variables.
    }
    \label{SI:unemploymentRegressionSimple}
\end{table}

\begin{table}[H]
    \centering
    \fontsize{4.5}{6}\selectfont
\begin{tabular}{l|lll|lll|lll|l}
\hline
\multicolumn{11}{c}{Dependent Variable: Log$_{10}$ Total Separations by State \& Month} \\ \hline
& \multicolumn{3}{c|}{Wave 1} & \multicolumn{3}{c|}{Wave 2} & \multicolumn{3}{c|}{Wave 3} & \\ 
                                                          Variable &        Model 1 &        Model 2 &        Model 3 &        Model 4 &        Model 5 &        Model 6 &        Model 7 &        Model 8 &        Model 9 &       Model 10 \\ \hline
          Acemoglu \& Autor$^{[14]}$ Computer Usage &               & -0.108$^{***}$ &                &               &               &                &                &                &                &         -0.006 \\
       Acemoglu \& Autor$^{[14]}$ Routine Cognitive &               &  0.176$^{***}$ &                &               &               &                &                &                &                &  0.385$^{***}$ \\
          Acemoglu \& Autor$^{[14]}$ Routine Manual &               & -0.194$^{***}$ &                &               &               &                &                &                &                & -0.618$^{***}$ \\
             ONET Education \% Workers w/ Bachelors &               &                & -0.143$^{***}$ &               &               &                &                &                &                & -0.914$^{***}$ \\
  Frey \& Osborne$^{[17]}$ Prob. of Computerisation &               &                &                &        -0.020 &               &                &                &                &                & -0.106$^{*  }$ \\
                  Arntz et al$^{[18]}$ Probability &               &                &                &               &        -0.018 &                &                &                &                & -0.196$^{***}$ \\
              ONET Automation Degree of Automation &               &                &                &               &               & -0.417$^{***}$ &                &                &                &         -0.096 \\
    Brynjolfsson et al$^{[23]}$ Suitability for ML &               &                &                &               &               &                & -0.236$^{***}$ &                &                & -0.772$^{***}$ \\
                 Felten et al$^{[24]}$ AI Exposure &               &                &                &               &               &                &                & -0.222$^{***}$ &                & -0.178$^{***}$ \\
                                Webb$^{[25]}$ \% AI &               &                &                &               &               &                &                &                &  0.111$^{** }$ &  1.345$^{***}$ \\
                             Webb$^{[25]}$ \% Robot &               &                &                &               &               &                &                &                &  0.210$^{***}$ &  0.983$^{***}$ \\
                          Webb$^{[25]}$ \% Software &               &                &                &               &               &                &                &                & -0.205$^{***}$ & -0.851$^{***}$ \\ \hline
                        $\log_{10}$ Wage Bill (\$) & 0.241$^{***}$ &  0.222$^{***}$ &  0.278$^{***}$ & 0.237$^{***}$ & 0.237$^{***}$ &  0.233$^{***}$ &  0.230$^{***}$ &  0.210$^{***}$ &  0.271$^{***}$ &  0.128$^{***}$ \\
                                         Year F.E. &           Yes &            Yes &            Yes &           Yes &           Yes &            Yes &            Yes &            Yes &            Yes &            Yes \\
                                        Month F.E. &           Yes &            Yes &            Yes &           Yes &           Yes &            Yes &            Yes &            Yes &            Yes &            Yes \\
                                             \hline
$R^2$ &         0.598 &          0.618 &          0.605 &         0.598 &         0.598 &          0.603 &          0.601 &          0.606 &          0.600 &          0.699 \\
                                        adj. $R^2$ &         0.597 &          0.616 &          0.603 &         0.597 &         0.597 &          0.601 &          0.599 &          0.604 &          0.599 &          0.698 \\
\hline \multicolumn{11}{c}{$p_{val}<0.1^*$, $p_{val}<0.01^{**}$, $p_{val}<0.001^{***}$} \\ \hline
\end{tabular}
    \caption{
        Multiple linear regression analysis of states' technology exposure and total unemployment rates each month according to LAUS.
        Data varies by state and month from January 2010 through 2020 for a total of 6,600 data points.
        All variables were centered and standardized before analysis thus eliminating units from the various AI exposure variables.
        We are unable to control for state fixed effects in this analysis, so we instead control for the size of each states' economy according to their total wage bill (i.e., total wages paid).
    }
    \label{SI:unemploymentRegression}
\end{table}

\subsection{State Job Separation Rates}
\label{SI:separations}
\begin{table}[H]
    \centering
    \fontsize{6.5}{7}\selectfont
    \begin{tabular}{l|ll|lll|lll}
    \hline
    \multicolumn{9}{c}{Dependent Variable: Log$_{10}$ Total Separations by State \& Month} \\ \hline
        & \multicolumn{2}{c|}{Wave 1} & \multicolumn{3}{c|}{Wave 2} & \multicolumn{3}{c}{Wave 3} \\ 
                                                          Variable &        Model 1 &        Model 2 &        Model 3 &        Model 4 &        Model 5 &        Model 6 &        Model 7 &        Model 8 \\ \hline
          Acemoglu \& Autor$^{[14]}$ Computer Usage &  0.077$^{***}$ &                &                &                &                &                &                &                \\
       Acemoglu \& Autor$^{[14]}$ Routine Cognitive & -0.143$^{***}$ &                &                &                &                &                &                &                \\
          Acemoglu \& Autor$^{[14]}$ Routine Manual &  0.399$^{***}$ &                &                &                &                &                &                &                \\
             ONET Education \% Workers w/ Bachelors &                & -0.212$^{***}$ &                &                &                &                &                &                \\
  Frey \& Osborne$^{[17]}$ Prob. of Computerisation &                &                & -0.088$^{***}$ &                &                &                &                &                \\
                  Arntz et al$^{[18]}$ Probability &                &                &                & -0.076$^{***}$ &                &                &                &                \\
              ONET Automation Degree of Automation &                &                &                &                & -0.157$^{***}$ &                &                &                \\
    Brynjolfsson et al$^{[23]}$ Suitability for ML &                &                &                &                &                & -0.170$^{***}$ &                &                \\
                 Felten et al$^{[24]}$ AI Exposure &                &                &                &                &                &                & -0.156$^{***}$ &                \\
                                Webb$^{[25]}$ \% AI &                &                &                &                &                &                &                & -0.257$^{***}$ \\
                             Webb$^{[25]}$ \% Robot &                &                &                &                &                &                &                &          0.023 \\
                          Webb$^{[25]}$ \% Software &                &                &                &                &                &                &                &  0.127$^{** }$ \\
                                             \hline
$R^2$ &          0.097 &          0.045 &          0.008 &          0.006 &          0.025 &          0.029 &          0.024 &          0.029 \\
                                        adj. $R^2$ &          0.096 &          0.045 &          0.008 &          0.006 &          0.024 &          0.029 &          0.024 &          0.028 \\
\hline \multicolumn{9}{c}{$p_{val}<0.1^*$, $p_{val}<0.01^{**}$, $p_{val}<0.001^{***}$} \\ \hline
\end{tabular}
    \caption{
        Linear regression analysis of states' technology exposure and total job separation rate according to JOLTS.
        Data varies by state, and month from January 2010 through 2020 for a total of 6,600 data points.
        All variables were centered and standardized before analysis thus eliminating units from the various exposure variables.
    }
    \label{SI:separationsRegressionSimple}
\end{table}

\begin{table}[H]
    \centering
    \fontsize{4.5}{6}\selectfont
\begin{tabular}{l|lll|lll|lll|l}
\hline
\multicolumn{11}{c}{Dependent Variable: Log$_{10}$ Total Separations by State \& Month} \\ \hline
& \multicolumn{3}{c|}{Wave 1} & \multicolumn{3}{c|}{Wave 2} & \multicolumn{3}{c|}{Wave 3} & \\ 
                                                          Variable &        Model 1 &        Model 2 &        Model 3 &        Model 4 &        Model 5 &        Model 6 &        Model 7 &        Model 8 &        Model 9 &       Model 10 \\ \hline
          Acemoglu \& Autor$^{[14]}$ Computer Usage &                &          0.003 &                &                &                &                &                &                &                &  0.406$^{***}$ \\
       Acemoglu \& Autor$^{[14]}$ Routine Cognitive &                &  0.068$^{***}$ &                &                &                &                &                &                &                &  0.105$^{***}$ \\
          Acemoglu \& Autor$^{[14]}$ Routine Manual &                &  0.116$^{***}$ &                &                &                &                &                &                &                & -0.072$^{*  }$ \\
             ONET Education \% Workers w/ Bachelors &                &                & -0.218$^{***}$ &                &                &                &                &                &                & -0.272$^{***}$ \\
  Frey \& Osborne$^{[17]}$ Prob. of Computerisation &                &                &                &  0.270$^{***}$ &                &                &                &                &                &  0.654$^{***}$ \\
                  Arntz et al$^{[18]}$ Probability &                &                &                &                &  0.222$^{***}$ &                &                &                &                & -0.247$^{***}$ \\
              ONET Automation Degree of Automation &                &                &                &                &                &  0.116$^{*  }$ &                &                &                &         -0.212 \\
    Brynjolfsson et al$^{[23]}$ Suitability for ML &                &                &                &                &                &                &  0.213$^{***}$ &                &                & -1.013$^{***}$ \\
                 Felten et al$^{[24]}$ AI Exposure &                &                &                &                &                &                &                &  0.352$^{***}$ &                &  0.518$^{***}$ \\
                                Webb$^{[25]}$ \% AI &                &                &                &                &                &                &                &                &         -0.001 &  0.221$^{** }$ \\
                             Webb$^{[25]}$ \% Robot &                &                &                &                &                &                &                &                &  0.280$^{***}$ &  0.672$^{***}$ \\
                          Webb$^{[25]}$ \% Software &                &                &                &                &                &                &                &                &         -0.002 & -0.320$^{***}$ \\ \hline
                        $\log_{10}$ Wage Bill (\$) & -0.259$^{***}$ & -0.220$^{***}$ & -0.203$^{***}$ & -0.197$^{***}$ & -0.202$^{***}$ & -0.257$^{***}$ & -0.249$^{***}$ & -0.210$^{***}$ & -0.189$^{***}$ & -0.159$^{***}$ \\
                                         Year F.E. &            Yes &            Yes &            Yes &            Yes &            Yes &            Yes &            Yes &            Yes &            Yes &            Yes \\
                                        Month F.E. &            Yes &            Yes &            Yes &            Yes &            Yes &            Yes &            Yes &            Yes &            Yes &            Yes \\
                                             \hline
$R^2$ &          0.379 &          0.401 &          0.394 &          0.402 &          0.394 &          0.379 &          0.381 &          0.398 &          0.404 &          0.457 \\
                                        adj. $R^2$ &          0.376 &          0.399 &          0.392 &          0.400 &          0.391 &          0.377 &          0.378 &          0.395 &          0.402 &          0.454 \\
\hline \multicolumn{11}{c}{$p_{val}<0.1^*$, $p_{val}<0.01^{**}$, $p_{val}<0.001^{***}$} \\ \hline
\end{tabular}
    \caption{
        Multiple linear regression analysis of states' technology exposure and total job separation rates each month according to JOLTS.
        Data varies by state and month from January 2010 through 2020 for a total of 6,600 data points.
        All variables were centered and standardized before analysis thus eliminating units from the various AI exposure variables.
        We are unable to control for state fixed effects in this analysis, so we instead control for the size of each states' economy according to their total wage bill (i.e., total wages paid).
    }
    \label{SI:joltsTotalSeparationRegression}
\end{table}

\begin{figure}[H]
    \centering
    \forloop{ct}{0}{\value{ct} < 8}{
        \includegraphics[width=.4\textwidth]{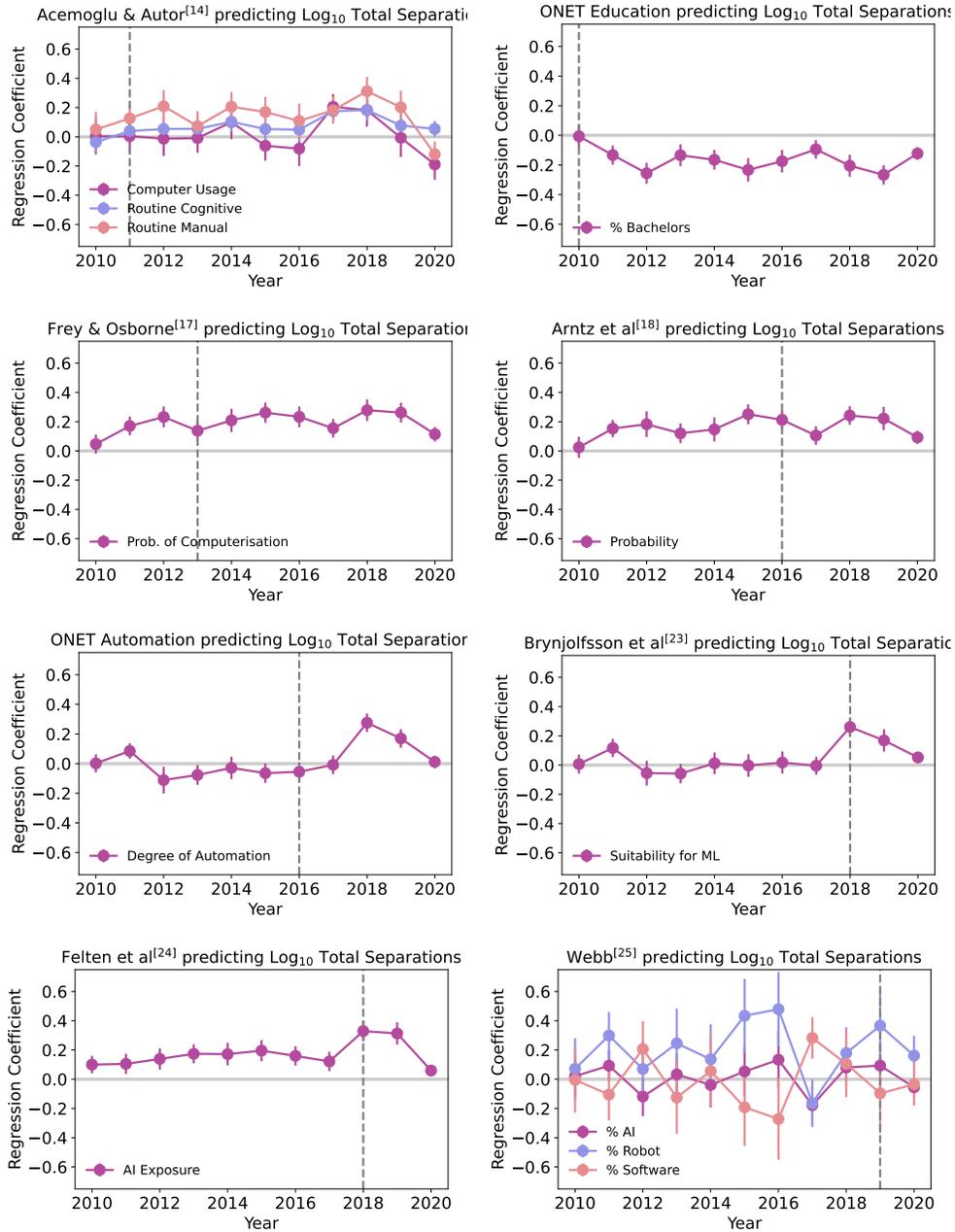}
    }
    \caption{
        Variables' regression coefficient estimate with monthly $\log_{10}$ total job separations as the dependent variable.
        Bars represent 95\% confidence intervals.
        Independent regressions with month fixed effects and were performed for each year while controlling for states' total wage bill (i.e., total wages paid).
        All variables were centered and standardized before analysis thus eliminating units from the various AI exposure variables.
    }
    \label{SI:totalSeparationsCoefOverTime}
\end{figure}

\begin{figure}[H]
    \centering
    \forloop{ct}{0}{\value{ct} < 8}{
        \includegraphics[width=.4\textwidth]{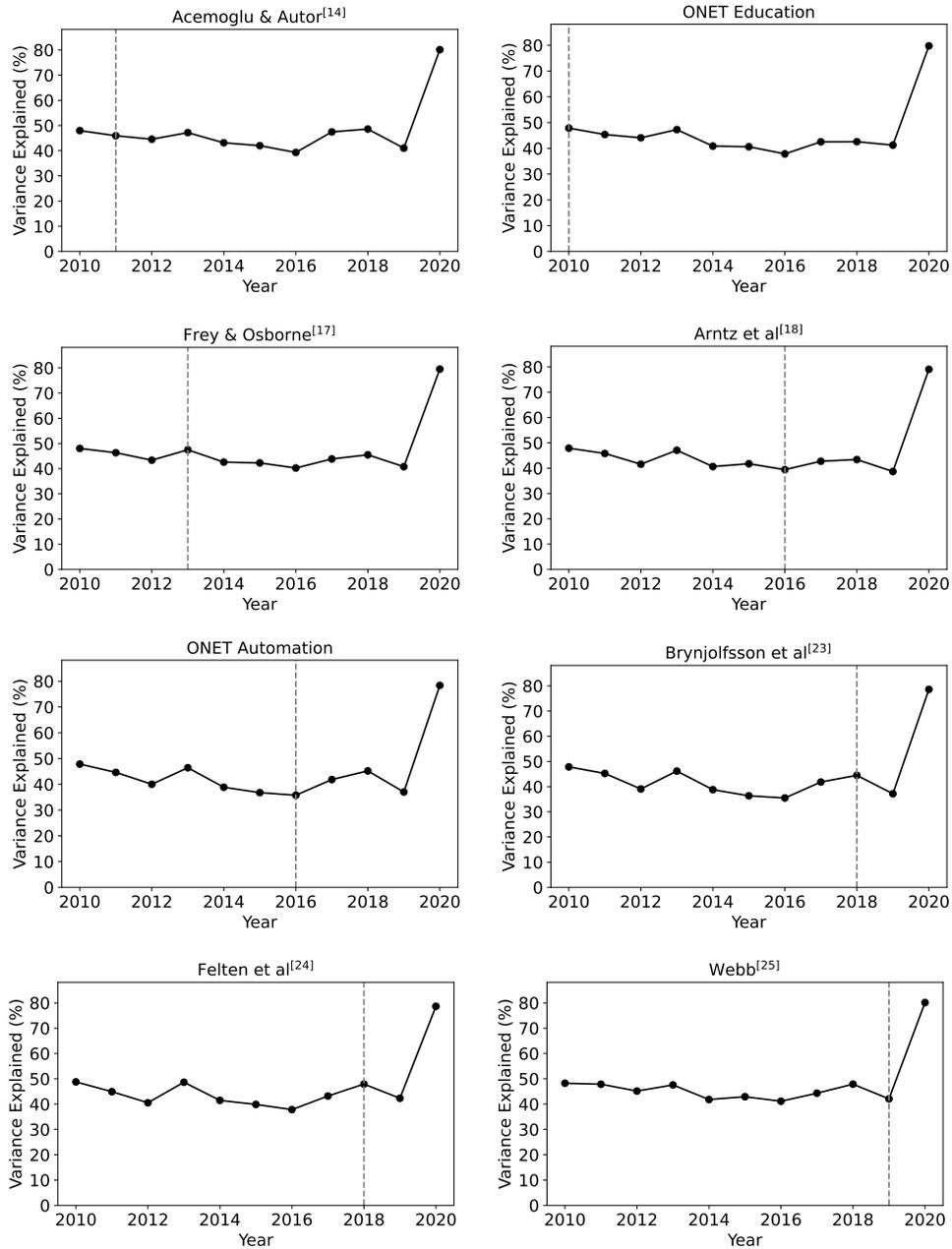}
    }
    \caption{
        Monthly $\log_{10}$ total job separations variance explained by each model.
        Independent regressions with month fixed effects and were performed for each year while controlling for states' total wage bill (i.e., total wages paid).
        All variables were centered and standardized before analysis thus eliminating units from the various AI exposure variables.
    }
    \label{SI:totalSeparationsROverTime}
\end{figure}

\begin{figure}[H]
    \centering
    \forloop{ct}{0}{\value{ct} < 8}{
        \includegraphics[width=.49\textwidth]{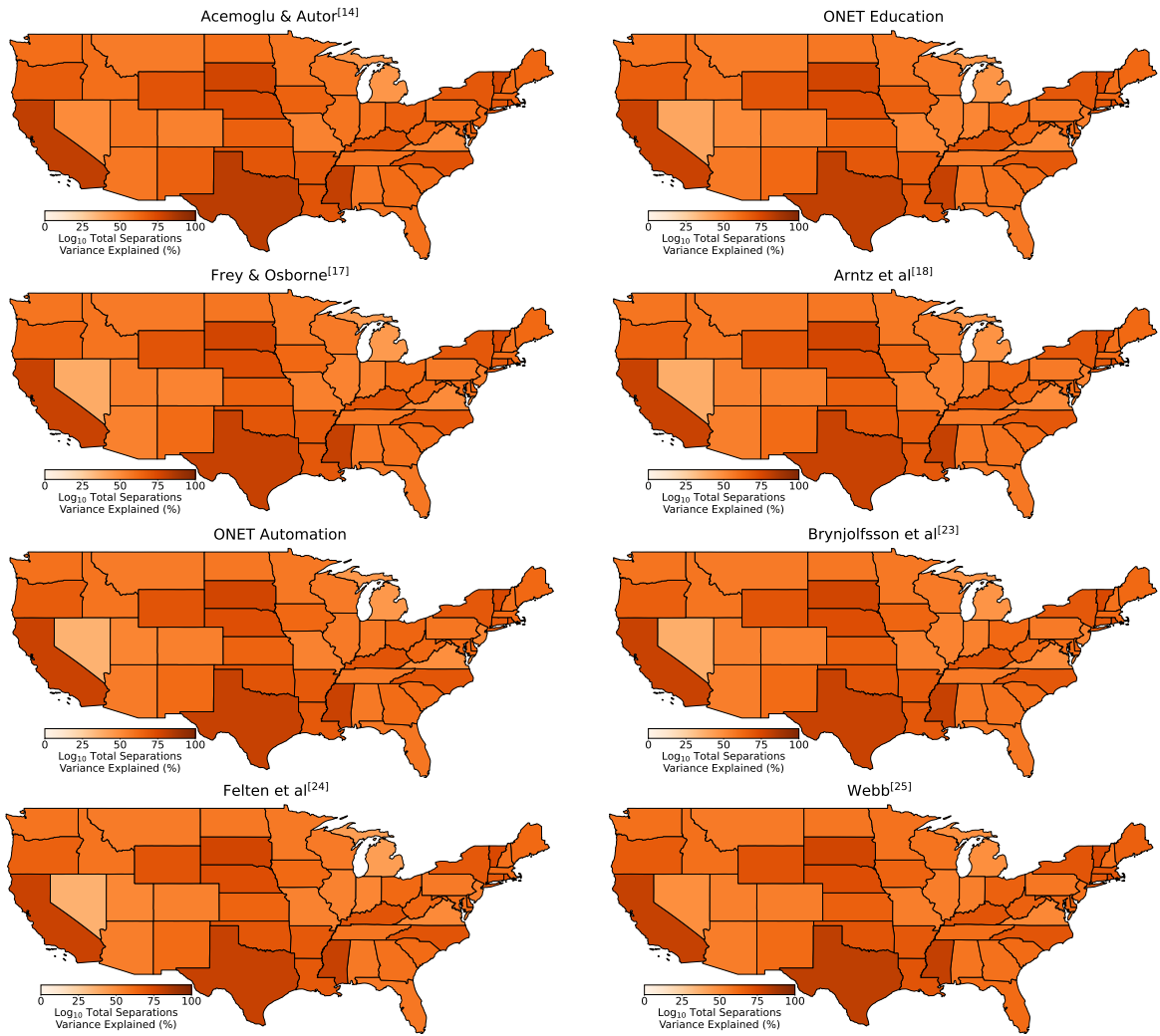}
    }
    \caption{
        Monthly $\log_{10}$ total separations variance explained by each model.
        Independent regressions with month fixed effects were performed for each state while controlling for states' total wage bill (i.e., total wages paid).
    }
    \label{SI:joltsTotalSeparationsRMap}
\end{figure}

\subsection{Within-Occupation Skill Change}
\label{SI:skillChangeSection}
\begin{figure}[H]
    \centering
    \includegraphics[width=.6\textwidth]{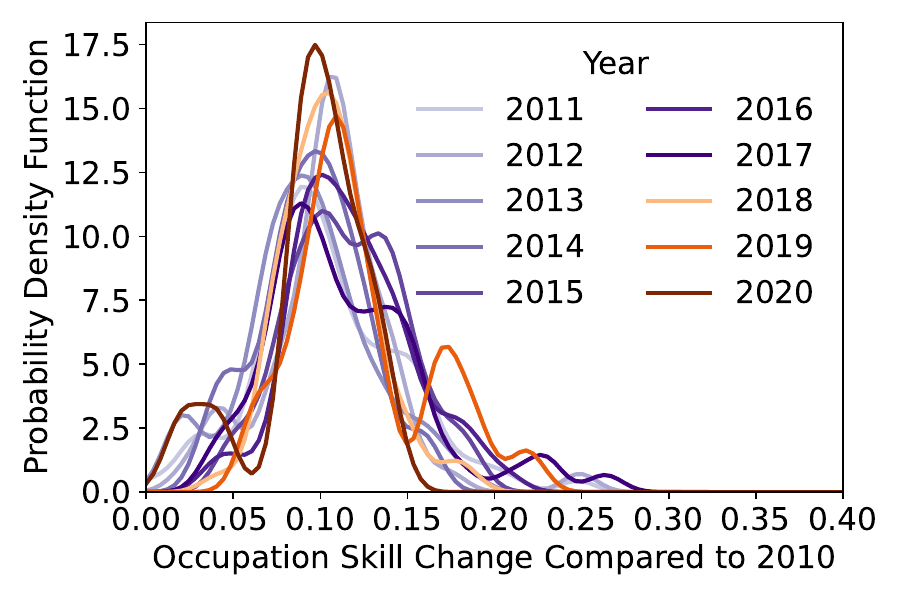}
    \caption{
        The annual distribution of within-occupation skill change compared to 2010.
        We consider only the occupations with skill profiles that were updated in each year.
        That is, each occupation contributes to the regression in only one year.
        We use 2011 through 2017 for the main analysis because the SOC taxonomy used to distinguish occupations was updated in 2018 thus introducing a potential confound to any longitudinal analysis.
    }
    \label{SI:skillChangeDist}
\end{figure}

We explore within-occupation changes to skill demand using annual occupation skill profiles from the BLS O*NET database.
After normalizing the varying Likert scales used across O*NET surveys, we obtain a skill vector for each occupation in each from 2010 through 2017.
Let $onet_{j_y,s}\in[0,1]$ represent the real-valued importance of skill $s$ to occupation $j$ in year $y$ such that $onet_{j_y,s}=1$ identifies an essential skill and $onet_{j_y,s}=0$ identifies an irrelevant skill.
Inspired by earlier work~\cite{gathmann2010general}, this data allows us to model each occupation as a vector of O*NET variables.
O*NET is updated annually, but each occupation is only updated every five years through a rolling survey.
That is, occupations are not updated each year. 
Therefore, for each six-digit Standard Occupation Classification (SOC) code $j$, we capture the within-occupation skill change in the year that $j$ was updated in O*NET and compare the updated skill profile to $j$'s skill requirements in 2010 according to
\begin{equation}
    \Delta skill(j_y,j_{2010})=1-\frac{\sum_{s\in S}\min(onet_{j_y,s},onet_{j_{2010},s})}{\sum_{s\in S}\max(onet_{j_y,s},onet_{j_{2010},s})}.
    \label{eq:jaccard}
\end{equation}
Notice that eq (\ref{eq:jaccard}) is one minus the Jaccard similarity and $\Delta skill \in [0,1]$.
We consider O*NET data only from 2010 through 2017 because the SOC taxonomy was updated in 2018.
Figure~\ref{SI:skillChangeDist} displays the distribution of skill change scores for the occupations that were updated in each year.
Although we provide distributions for 2018, 2019, and 2020 in Fig.~\ref{SI:skillChangeDist}, all other analyses of skill change in this study are restricted to O*NET data from 2010 through 2017 because the SOC taxonomy was updated from 2018 on-wards (i.e., the list of occupation codes changed) which introduces the potential for skill changes to result from the merging or separation of 2010 SOC codes.

\begin{table}[H]
    \centering
    \fontsize{6.5}{6}\selectfont
\begin{tabular}{l|ll|lll|lll}
\hline\multicolumn{9}{c}{Dependent Variable: Within-Occupation Skill Change compared to 2010 by Detailed Occupation} \\ \hline
    & \multicolumn{2}{c|}{Wave 1} & \multicolumn{3}{c|}{Wave 2} & \multicolumn{3}{c}{Wave 3} \\ 
    Variable &        Model 1 &        Model 2 &       Model 3 &       Model 4 & Model 5 &       Model 6 &        Model 7 &        Model 8 \\ \hline
          Acemoglu \& Autor$^{[14]}$ Computer Usage & -0.275$^{***}$ &                &               &               &         &               &                &                \\
       Acemoglu \& Autor$^{[14]}$ Routine Cognitive &          0.062 &                &               &               &         &               &                &                \\
          Acemoglu \& Autor$^{[14]}$ Routine Manual &          0.069 &                &               &               &         &               &                &                \\
             ONET Education \% Workers w/ Bachelors &                & -0.216$^{***}$ &               &               &         &               &                &                \\
  Frey \& Osborne$^{[17]}$ Prob. of Computerisation &                &                & 0.345$^{***}$ &               &         &               &                &                \\
                  Arntz et al$^{[18]}$ Probability &                &                &               & 0.245$^{***}$ &         &               &                &                \\
              ONET Automation Degree of Automation &                &                &               &               &   0.050 &               &                &                \\
    Brynjolfsson et al$^{[23]}$ Suitability for ML &                &                &               &               &         & 0.363$^{***}$ &                &                \\
                 Felten et al$^{[24]}$ AI Exposure &                &                &               &               &         &               & -0.256$^{***}$ &                \\
                                Webb$^{[25]}$ \% AI &                &                &               &               &         &               &                & -0.203$^{***}$ \\
                             Webb$^{[25]}$ \% Robot &                &                &               &               &         &               &                &  0.261$^{***}$ \\
                          Webb$^{[25]}$ \% Software &                &                &               &               &         &               &                &          0.016 \\
                                         \hline Year F.E. &             No &             No &            No &            No &      No &            No &             No &             No \\
                                    Major SOC F.E. &             No &             No &            No &            No &      No &            No &             No &             No \\
                                             \hline
$R^2$ &          0.108 &          0.047 &         0.119 &         0.060 &   0.002 &         0.132 &          0.065 &          0.081 \\
                                        adj. $R^2$ &          0.103 &          0.045 &         0.117 &         0.058 &   0.001 &         0.130 &          0.064 &          0.076 \\
\hline \multicolumn{9}{c}{$p_{val}<0.1^*$, $p_{val}<0.01^{**}$, $p_{val}<0.001^{***}$} \\ \hline
\end{tabular}
    \caption{
        Multiple linear regression analysis of within-occupation skill change compared to 2010 O*NET skill profiles.
        Data varies by six-digit SOC code from 2011 through 2017 for a total of 479 data points.
        All variables were centered and standardized before analysis thus eliminating potentially different units from the technology exposure scores.
    }
    \label{SI:skillChangeSimpleRegression}
\end{table}

\begin{table}[H]
    \centering
    \fontsize{5}{6}\selectfont
\begin{tabular}{l|lll|lll|lll|l}
\hline\multicolumn{11}{c}{Dependent Variable: Within-Occupation Skill Change compared to 2010 by Detailed Occupation} \\ \hline
    & \multicolumn{3}{c|}{Wave 1} & \multicolumn{3}{c|}{Wave 2} & \multicolumn{3}{c|}{Wave 3} & \\
                                                Variable & Model 1 &        Model 2 &        Model 3 &       Model 4 & Model 5 & Model 6 &       Model 7 &        Model 8 &       Model 9 &       Model 10 \\ \hline
          Acemoglu \& Autor$^{[14]}$ Computer Usage &         & -0.214$^{***}$ &                &               &         &         &               &                &               & -0.118$^{*  }$ \\
       Acemoglu \& Autor$^{[14]}$ Routine Cognitive &         &         -0.035 &                &               &         &         &               &                &               &         -0.079 \\
          Acemoglu \& Autor$^{[14]}$ Routine Manual &         &          0.096 &                &               &         &         &               &                &               &          0.034 \\
             ONET Education \% Workers w/ Bachelors &         &                & -0.103$^{*  }$ &               &         &         &               &                &               &          0.051 \\
  Frey \& Osborne$^{[17]}$ Prob. of Computerisation &         &                &                & 0.220$^{***}$ &         &         &               &                &               &          0.075 \\
                  Arntz et al$^{[18]}$ Probability &         &                &                &               &   0.075 &         &               &                &               &          0.032 \\
              ONET Automation Degree of Automation &         &                &                &               &         &  -0.057 &               &                &               &         -0.066 \\
    Brynjolfsson et al$^{[23]}$ Suitability for ML &         &                &                &               &         &         & 0.390$^{***}$ &                &               &  0.330$^{***}$ \\
                 Felten et al$^{[24]}$ AI Exposure &         &                &                &               &         &         &               & -0.343$^{***}$ &               & -0.269$^{***}$ \\
                                Webb$^{[25]}$ \% AI &         &                &                &               &         &         &               &                &        -0.107 &          0.008 \\
                             Webb$^{[25]}$ \% Robot &         &                &                &               &         &         &               &                & 0.242$^{** }$ &          0.120 \\
                          Webb$^{[25]}$ \% Software &         &                &                &               &         &         &               &                &        -0.022 &         -0.027 \\
                                         \hline Year F.E. &     Yes &            Yes &            Yes &           Yes &     Yes &     Yes &           Yes &            Yes &           Yes &            Yes \\
                                    Major SOC F.E. &     Yes &            Yes &            Yes &           Yes &     Yes &     Yes &           Yes &            Yes &           Yes &            Yes \\
                                             \hline
$R^2$ &   0.251 &          0.274 &          0.257 &         0.272 &   0.255 &   0.253 &         0.292 &          0.306 &         0.268 &          0.357 \\
                                        adj. $R^2$ &   0.207 &          0.228 &          0.212 &         0.228 &   0.209 &   0.208 &         0.249 &          0.264 &         0.221 &          0.304 \\
\hline \multicolumn{11}{c}{$p_{val}<0.1^*$, $p_{val}<0.01^{**}$, $p_{val}<0.001^{***}$} \\ \hline
\end{tabular}
    \caption{
        Multiple linear regression analysis of within-occupation skill change compared to 2010 O*NET skill profiles.
        Data varies by six-digit SOC code from 2011 through 2017 for a total of 479 data points.
        All variables were centered and standardized before analysis, thus eliminating potentially different units from the technology exposure scores.
    }
    \label{SI:skillChangeRegression}
\end{table}

\end{document}